%% file: ms.tex
\documentclass{emulateapj}

\usepackage{verbatim}
\usepackage{graphicx}
\usepackage{multirow}
\usepackage{longtable}
\usepackage{verbatim}
\usepackage{setspace}

\begin{document}

\title{Evolutionary Channels for the Formation of Double Neutron Stars}
\author{Jeff J. Andrews\altaffilmark{1}, W. M. Farr\altaffilmark{2,3}, V. Kalogera\altaffilmark{2}, B. Willems\altaffilmark{4}}

\altaffiltext{1}{Columbia University, 550 West 120th St., New York, NY 10027}
\altaffiltext{2}{CIERA (Center for Interdisciplinary Exploration and Research in Astrophysics) \& Dept.\ of Physics and Astronomy, Northwestern University, 2145 Sheridan Rd, Evanston, IL 60208}
\altaffiltext{3}{School of Physcis and Astronomy, University of Birmingham, Birmingham, B15 2TT, United Kingdom}
\altaffiltext{4}{Atipa Technologies, 4921 Legends Drive, Lawrence, KS 66049 (current address) }

\begin{abstract}

We analyze binary population models of double-neutron stars and compare results to the accurately measured orbital periods and eccentricities of the eight known such systems in our Galaxy. In contrast to past similar studies, we especially focus on the dominant evolutionary channels (we identify three); for the first time, we use a detailed understanding of the evolutionary history of three double neutron stars as actual constraints on the population models. We find that the evolutionary constraints derived from the double pulsar are particularly tight, and less than half of the examined models survive the full set of constraints. The top-likelihood surviving models yield constraints on the key binary evolution parameters, but most interestingly reveal (i) the need for electron-capture supernovae from relatively low-mass degenerate, progenitor cores, and (ii) the most likely evolutionary paths for the rest of the known double neutron stars. In particular, we find that J1913+16 likely went through a phase of Case BB mass transfer, and J1906+0746 and J1756$-$2251 are consistent with having been formed in electron-capture supernovae.

\end{abstract}

\section{Introduction}
\label{intro}

Since the discovery of the first double neutron star (DNS) system
\citep{HT75} almost 40 years ago, astronomers have discovered only 8
more DNS among more than 2000 pulsars.  As exotic endpoints of binary
evolution, DNS form through a wide variety of binary phases, including
tides, strong stellar winds, two supernovae (SN), multiple phases of
mass transfer (stable and unstable), and orbital decay due to
gravitational wave radiation \citep{vdHH72,FvdH75,DT81,BvdH91,TY93}.  Binary disruption due to
a SN or coalescence during dynamically unstable mass transfer and
common-envelope (CE) evolution are both highly likely, yet all
observed DNS have survived these phases. Without two supernovae and CE
evolution, the observed DNS in their tight orbits could not have
formed.

Our current understanding of the full CE evolution and the final
outcome for the binaries going through it is incomplete, and
consequently CE modeling for binary populations has been limited to
relying on simple parameterizations of the problem which involve the
amount of mass and energy or angular momentum lost in this
non-conservative process
\citep{W84,LS88,dK90,IL93,NVYP00,W08,TR10}. Although such
parameterizations are of limited use in terms of understanding the
physics of this phase, they have allowed considerable progress in
understanding the formation and evolution of interacting binary
populations with compact objects (white dwarfs, neutron stars, black
holes). Thus there is always interest in constraining them empirically
using observed systems and populations in the Milky Way and other
galaxies. Recently, more physically motivated treatments of the CE
phase have been put forward \citep[][for a review]{DT10,GHW10,WI11,IC11,dM11,IJC13}. These represent important advancements, however they require the use
of a stellar structure and evolution code as part of the modeling, and
therefore they are not applicable to the current tools available for
population syntheses (which adopt fitting formulae based on single star
evolution).

Binaries have a significant chance of getting disrupted when one of
the stars goes through a SN, either because of the associated mass
loss and/or because of asymmetries in the underlying mechanism
imparting a ``natal kick'' to the newborn NS of hundreds of km/s.
Based on observations of proper motions of single pulsars, not only do
\citet{Hobbs05} derive a Maxwellian kick distribution with a
dispersion velocity of 265 km/s, but they argue in favor of an absence
of pulsars with very low proper motions.  However an early population
synthesis study by \citet{PZY98} argued that the formation of the
observed systems requires a distribution with a substantial fraction
of NS born with low velocity kicks.  \citet{PRPS02} arrived at the
same conclusion, finding that low kick velocities are necessary to
explain the observed population of Be/X-ray binaries, and the presence
of large numbers of neutron stars in Galactic globular clusters
\citep{PRP02}. In search of a physical explanation for how some
neutron stars may form with small natal kicks, \citet{Pod04} drew
renewed attention to a core-collapse mechanism known as an
electron-capture supernova (ECS).  Originally proposed by
\citet{MN80}, an ECS is a low energy SN occurring when an ONe
degenerate core-collapses due to electron captures onto $^{24}$Mg and
$^{20}$Ne \citep{N84,MN87,N87}.  \citet{Pod04} postulated that this
channel was only accessible to binary stars that had lost their
envelopes in a mass transfer phase, thereby limiting the second dredge
up phase.  The remnant NS is predicted to receive a kick of
$\lesssim50-100$\,km/s. Very soon after the discovery of the double
pulsar \citep{BDP03}, analysis of its properties led a number of
groups \citep{WK04,DvdH04,PS05,S06,WKF06,D10,WWK10,FSK13} to argue
(albeit for different reasons and at different levels of quantitative
analysis) in favor of low-velocity natal kicks being involved in the
formation of the second pulsar. In a qualitative analysis
\citet{vdH07} expanded the picture arguing that through both
mechanisms, the standard Fe core-collapse and the ECS, the entire
population of DNS could be formed. Most recently \citet{WWK10}
analyzed all individual systems in view of all current observational
properties (binary and kinematic) and concluded that some observed DNS
must have received high natal kicks and other systems require low
natal kicks, also indicating that both mechanisms are needed.

\begin{deluxetable*}{lccccccc}

\tablecaption{Parameters of the Known Double Neutron Stars \label{tab:known_DNS}}
\tablehead{
\colhead{DNS} &
\colhead{$e$\tablenotemark{d}} &
\colhead{$P$\tablenotemark{d}} &
\colhead{$P_s$} &
\colhead{Pulsar Mass} &
\colhead{Companion Mass} & 
\colhead{Evol. Constraints\tablenotemark{e}} &
\colhead{References} \\
& &
\colhead{[d]} &
\colhead{[ms]} &
\colhead{$M_{\odot}$} &
\colhead{$M_{\odot}$} & & 
}
\startdata
	B1534+12 & 0.274 & 0.421 & 37.9 & 1.3332(10) & 1.3452(10) & Channel II & 1 \\
	B1913+16 & 0.617 & 0.323 & 59.0 & 1.4408(3) & 1.3873(3) & Channel I or II & 2, 3 \\
	J0737$-$3039 & 0.088 & 0.102 & 22.7 & 1.337(5) & 1.250(5) & Channel III & 4 \\
	J1518+4904 & 0.249 & 8.634 & 40.9 & 1.56$^{+0.13}_{-0.45}$ & 1.05$^{+0.45}_{-0.11}$ & & 5 \\
	J1756$-$2251 & 0.181 & 0.320 & 28.5 & 1.341(7) & 1.230(7) &  & 6 \\
	J1811$-$1736 & 0.828 & 18.779 & 104.2 & 1.62$^{+0.22}_{-0.55}$ & 1.11$^{+0.53}_{-0.15}$ & & 7 \\
	J1829+2456 & 0.139 & 1.176 & 41.0 & 1.14$^{+0.28}_{-0.48}$ & 1.36$^{+0.50}_{-0.17}$ &  & 8 \\
	J1906+0746\tablenotemark{a} & 0.085 & 0.166 & 144.1 & 1.248(18) & 1.365(18) & & 9, 10 \\
	\hline
	J1753$-$2240\tablenotemark{b} & 0.304 & 13.638 & 95.1&  &  &  & 11 \\
	B2127+11C\tablenotemark{c} & 0.680 & 0.335 & 30.5 & 1.35(4) & 1.36(4) &  & 12
\enddata
\tablerefs{1 -- \citet{W91}, 2 -- \citet{HT75}, 3 -- \citet{TW82}, 4 -- \citet{BDP03}, 5 -- \citet{J08}, 6 -- \citet{FSK14}, 7 -- \citet{C04}, 8 -- \citet{Ch04}, 9 -- \citet{L06}, 10 -- \citet{K08}, 11 -- \citet{KKL09}, 12 -- \citet{J06}}
\tablenotetext{a}{The observed pulsar in J1906+0746 is the unrecycled secondary.} 
\tablenotetext{b}{J1753$-$2240 is unconfirmed as a DNS, we therefore exclude it from our analysis.  See \S \ref{indiv_systems}}
\tablenotetext{c}{B2127+11C is contained within the globular cluster M15 and is suspected to have formed dynamically, therefore we exclude it from our analysis.}
\tablenotetext{d}{Values for eccentricity and orbital period are rounded to 3 decimal places.}
\tablenotetext{e}{We only use conservative evolutionary constraints.  Most systems do not have enough information to place any constraints on the evolution.  See \S \ref{evol_constrain} and references for a more complete explanation.}
\end{deluxetable*}

In the study presented here we develop population models of DNS
formation and evolution and we contrast our results against DNS
measurements of orbital period $P$ and eccentricity $e$ (only).
We complement upon the work of previous population synthesis studies of DNS \citep{KHB08,KHB10,OBG11,DBF12} by adding the 
 consideration of any constraints on the evolutionary channel
each of the observed systems has followed; we use a Bayesian analysis
to quantitatively assess the results. The last analysis to examine
whether measured $P$ and $e$ values are in agreement with DNS
population modeling was presented by \citet{D10}, who focused
primarily on the formation of J0737$-$3039, with no quantitative
statistical analysis. 

We undertake this study to address the following specific questions:
(i) are there current DNS population models that are consistent with
DNS $P$ and $e$ measurements? (ii) can such consistent models also
account for the strong evolutionary constraints that are currently
available for a few of the observed systems? (iii) if yes, then how do
these select models constrain binary evolution processes and
parameters; which processes and parameters are not important? (iv) can
we use the select models to uncover the evolutionary history of the
rest of the observed DNS systems? We choose to restrict ourselves to
the $P$ and $e$ measurements for a number of reasons: they are the
best measured DNS properties (without any indirect inference
involved), they are highly correlated making it easy to evaluate
the models in a 2-dimensional plane, and population models make
reliable predictions for these properties, unlike others (e.g.,
rates).  Our additional consideration of evolutionary constraints for
specific systems introduces a new element, never used before in DNS
population synthesis.  In \S \ref{method} we describe our modeling
method followed by our quantitative analysis method in \S
\ref{model_analy}.  In \S \ref{results}, we present our results and
discuss them in \S \ref{discussion}, ending with a summary in \S
\ref{sum}.

\section{Modeling Methods and Simulations}
\label{method}

\subsection{StarTrack Population Synthesis Code and Models in This Study}
\label{startrack}

To simulate DNS formation, we adopt the population synthesis code
\texttt{StarTrack} \citep{BKR08}. We use a reference model with a
given set of parameters, and we then vary a large number of them from
the reference assumptions to assess their effect on DNS properties and
formation channels.

\texttt{StarTrack} adapts the fitting formulae by \citet{HPT00} for
single star modeling (time evolution of key macroscopic physical
properties) and it modifies them to account for the effects of mass
transfer to allow for modeling of binary evolution. The mass-transfer
treatment (outside of the case of CE phases) has been tested against
and calibrated to mass-transfer sequences \citep[see][and references
  therein]{BKR08}. This code has been used for a wide range of studies
and applications involving modeling of interacting binaries with all
types of compact objects, and has been compared to and demonstrated
agreement with observations of such binaries both in our Galaxy and in
external galaxies since the early 2000's when the first version of the
code was produced. It has also been maintained and kept up to date as
our understanding of compact object formation and stellar winds has
evolved over the years. In particular, updated prescriptions for
massive stellar winds and their dependence on metallicity have been
included \citep{BBF10}. Although a recent revision of the code uses a
compact object remnant mass function physically motivated by
calculations of the supernova mechanism \citep{FBW12}, our results
rely on the remnant mass function calculated by \citet{TWW96}.  An
additional recent parametrization of the envelope binding energy
parameter $\lambda$, a measure of the central concentration of a
stellar envelope which is relevant for CE calculations,
was added to \texttt{Startrack} after our simulations were completed
\citep{DBF12} and is not included in the version of the code used
here. 

%Although we assume here that $\lambda=1.0$, \citep{DBF12} find that for 10 M$_{\odot}$ giant stars (representative of a NS progenitor), lambda values range between $\sim0.4 - 1.2$, depending on its evolutionary state and radius at the time of Roche lobe overflow. For values of $\lambda$ less than unity, tighter binaries will be produced than what is represented by our code. 

Initial conditions for each population model are as follows: the
primary star in the binary is formed on the main sequence with a mass
drawn from a power law distribution $\sim M^{-2.7}$ \citep{S86}, while
the secondary is determined based on the mass ratio value drawn
randomly from a flat distribution between 0 and 1; the initial orbital
separation is logarithmically flat and the initial eccentricity
follows the dynamical-equilibrium distribution $\sim 2e$.  Allowing
for wind mass loss, the primary star is then followed through the
Hertzsprung gap (HG), red giant branch, He-main sequence, and
asymptotic giant branch (AGB).  If the primary is massive enough, it
will collapse through a SN, leaving a compact remnant.  Assuming the
binary is not disrupted during the first SN, the secondary evolves
through its lifetime, with possible interruptions by mass transfer
sequences and also driven by wind mass loss and/or tides.  Once it has
evolved completely, if massive enough, the secondary will also leave a
compact remnant as the product of a SN.  Asymmetries during the SN
impart an (assumed instantaneous with respect to the orbital period)
kick to the neutron star or black hole, altering the shape of the
binary orbit, following the equations of \citet{K96,K00}.  The
formation of a DNS requires that both stars have pre-SN masses within
a specific range, such that the binary neither disrupts due to the SN
kick nor does it merge prior to the second SN.  The particular
variables and prescriptions used are given in \S \ref{ref_model}.  The
DNS population is given a constant birth rate over the past 10 Gyrs.
After the second SN, the orbital evolution due to gravitational
radiation is followed using the equations of \citet{P64} until the
current epoch.

\subsubsection{Reference Model}
\label{ref_model}

Our reference model (Model 1) adopts most of the the standard values
suggested in \citet{BKR08}. The few modifications are described below.

An Fe core-collapse supernova occurs when a star forms a
non-degenerate CO core massive enough to ignite stable nuclear
burning, creating progressively heavier elements until the degeneracy
pressure is overcome and the Fe core burns explosively. If instead the
CO core is formed partially degenerate and reaches a critical mass of
1.08 $M_{\odot}$, stable nuclear burning will ensue, creating a
degenerate ONe core. We assume that the entire CO core is burned to form the ONe core and approximate that this grows at the same rate that the CO core would. A resulting ONe core more massive than 1.38
$M_{\odot}$ is thought to collapse in an electron capture supernova (ECS) while less massive cores form
ONe white dwarfs (WD). In our code we vary the range within which the He core mass
forms a partially degenerate CO core ($M_{\rm He,deg}$). In our
reference model, we set this range to 2.0 $\le$ $M_{\rm
  He,deg}$/$M_{\odot}$ $\le$ 2.5.

To stars undergoing an Fe core-collapse SN, we apply kick velocities randomly chosen from a Maxwellian distribution with a dispersion velocity of 300 km/s, in agreement with observations of the space velocities of single pulsars \citep{Hobbs05}. However \citet{KJH06} find that the neutrino heating mechanism occurs faster and produces a smaller ejecta mass in an ECS, and they conclude that ECS events impart a smaller kick velocity to collapsing ONe degenerate cores ($\lesssim 100$ km/s). In practice these specific boundary values on the CO core, as well as its progenitor He-rich degenerate core, are quite uncertain. Therefore in our code, we apply a kick decreased by a factor of 10 to a star collapsing in an ECS compared with the standard Fe-core case. 

The mass transfer parameters on which we focus in this analysis
primarily deal with CE evolution.  These phases are treated
using the equations provided by \citet{W84} and \citet{dK90}.  In this
prescription, the efficiency of orbital energy loss during a CE phase
is determined by the efficiency parameter, $\alpha_{CE}$, and the
binding energy of an envelope is parametrized with $\lambda$.  
Since the two parameters are degenerate in the energy conservation formalism for CE evolution, in our models we vary the product, $\alpha_{\rm CE}\lambda$. We set $\alpha_{\rm CE}\lambda$ to be 0.5 for the reference model.  
%Although we do not formally vary $\lambda$, when we discuss varying the value of $\alpha_{\rm CE}$, in practice we are varying the product, $\alpha_{\rm CE}\lambda$, since the two parameters are degenerate in the energy conservation formalism for CE evolution. 
We do not attempt to calculate the mass accreted within
a CE, instead limiting the mass accreted to a random amount between 0.05-0.1
$M_{\odot}$.  The fraction of mass lost that is accreted onto a main
sequence star during stable Roche lobe overflow is set at 0.5 for the
reference model. Stellar evolution codes show that HG stars lack a
steep entropy gradient across the core-envelope boundary \citep{IT04},
and numerical results indicate that a HG star will merge with its
companion upon entering a CE \citep{TS00}. In the reference case, all
CE with HG donors result in a merger.

We set the maximum NS mass to 2.5 $M_{\odot}$, where NS that become
too massive collapse as black holes. See \citet{BKR08} for additional
parameters dealing with the calculation of winds, tides, and stellar
structure in the reference model.

\subsubsection{Varied Parameters}
\label{model_params}

\setlength{\tabcolsep}{3pt}

\begin{deluxetable}{cllllll}
\tablecaption{Top Models \label{tab:top_models}}
\tablehead{
\colhead{Model} & 
\colhead{$ \alpha_{\rm CE}\lambda$} & 
\colhead{Kick\tablenotemark{a}} & 
\colhead{Notes\tablenotemark{b}} & 
\colhead{log($ \Lambda$)} & 
\colhead{log($ \Lambda_{\rm evol}$)} & 
\colhead{rank\tablenotemark{c}} \\ 
 & & \colhead{[km/s]} & & & &
}
\startdata
1 & 0.5 & 300 &  & -23.2 & -23.8 & 35 \\ 
2 & 0.25 & 300 &  & -25.2 & -26.4 & 75 \\ 
3 & 0.3 & 300 &  & -23.5 & -24.4 & 54 \\ 
4 & 1.0 & 300 &  & -23.0 & -25.2 & 64 \\ 
5 & 0.5 & 300 & HCE & -23.4 & -24.2 & 46 \\ 
6 & 0.5 & 300 & HG in CE & -22.8 & -23.4 & 16 \\ 
7 & 0.5 & 300 & HCE, HG in CE & -22.8 & -23.3 & 11 \\ 
8 & 0.5 & 50 &  & -21.9 & -23.4 & 14 \\ 
9 & 0.5 & 150 &  & -22.4 & -23.5 & 20 
\enddata 

\tablenotetext{a}{The kick velocity applied to a NS born in an Fe core-collapse SN is drawn randomly from a Maxwellian distribution with this dispersion velocity.} 
\tablenotetext{b}{HCE: NS accrete hypercritically in the CE; HG in CE: HG stars are allowed to survive a CE.} 
\tablenotetext{c}{The rank of each model when ordered by $\Lambda_{\rm evol}$.}
\tablecomments{These models all have reference values (found in \S \ref{ref_model}) for parameters unlisted here.} 
\end{deluxetable}

\setlength{\tabcolsep}{6pt}

Overall, we tested 155 separate models, each of which is listed in
Tables \ref{tab:top_models} and \ref{tab:models_add}.

We explore the effects of different Fe core-collapse supernova kick
velocities by testing Maxwellian distributions with smaller dispersion
velocities of 150 km/s, 50 km/s, and a model with no kicks. We also
test models in which the ECS kick velocity is varied, including a model in which all SN kicks (Fe core-collapse and ECS) have a velocity dispersion of 50 km/s. We test models
with the following ranges for $M_{\rm He,deg}$ which form a partially
degenerate CO core: 1.3-2.25, 1.7-1.9, 1.8-2.1, 1.83-2.25, 1.83-2.5,
2.0-2.5, 2.2-3.0, 2.5-2.7 and 1.66-3.24 $M_{\odot}$.  These ranges of
masses are varied to include lower, higher, and wider ranges than the
reference range.  In dealing with the CE, we explore the parameter
space of $\alpha_{\rm CE}\lambda$, testing both higher and lower values: 0.01,
0.05, 0.1, 0.2, 0.25, 0.3, and 1.0. We do not test values of
$\alpha_{\rm CE}\lambda$ $>$ 1.0 corresponding to an efficiency greater than
100\% as this requires an additional energy source.  We also test an
alternative prescription, formulated by \citet{NVYP00} in which
angular momentum, not energy, is parametrized \citep[see
  also][]{NT05}.  We further test models in which we attempt to calculate the mass accreted by a NS in a CE, set to be half the Bondi-Hoyle accretion rate. Finally, we test models allowing HG donors to survive a CE.

The maximum NS mass is set at 2.5 $M_{\odot}$, but
we test models lowering the limit to 2.0 $M_{\odot}$, which becomes
relevant in models allowing hypercritical accretion in the CE.  We
alter the fraction of mass lost in a stable semi-detached system by a
non-compact accretor between 0, 0.1, 0.3, 1.0, and the default value
of 0.5.  Other variations from the reference model include decreasing
the wind strengths, decreasing the effectiveness of tides, altering
the mass below which the helium envelope becomes convective, using an
alternative initial mass function for the secondary to create binaries
with similar initial masses, altering the mass-radius relation for
helium stars, increasing the specific angular momentum of matter, and
various combinations of these parameters.

We also adopt a couple more assumptions which we do not vary in our set of models. One such variable is the NS birth mass (1.35 $M_{\odot}$ for Fe core-collapse SN and 1.26 M$_{\odot}$ for an ECS). We make this assumption based on observed masses of NS, and we expect that slightly altering the birth mass of NS will not significantly affect the orbital periods and eccentricities of the DNS produced \citep[see][and references therein]{OPN12}. We also do not vary the star formation rate. Although this is a simple approximation, the star formation rate probably does not vary by much more than a factor of 2 over the past 10 Gyr \citep{RpSM00}.

\section{Model Analysis Methodology}
\label{model_analy}

We analyze our simulation results in the context of the DNS orbital
period-eccentricity plane, and their evolutionary constraints.  In
Figure \ref{stnd_mod} the distribution of the entire DNS population
for our reference model is shown in the first panel.  The eight
observed systems are overlaid as diamonds on top of the distribution.
The 68.3\%, 95.45\%, 99.7\%, and 99.994\% confidence levels of the DNS
population are displayed by contours in Figure \ref{stnd_mod}.  The
characteristic band, running from low eccentricity to high
eccentricity is due to the finite impulse of energy and angular
momentum by a kick imparted to the second NS at its birth.  Inspiral
and circularization due to the emission of gravitational radiation
affect systems with short orbital periods and large eccentricities.
This band of systems qualitatively matches the distributions found by
previous authors \citep{PZY98,BB99,D10,KHB10}.  We expand on the work of
previous authors by including the three additional plots in Figure
\ref{stnd_mod}.  These plots break down the entire population into
three parts based on their evolutionary channels described in the \S
\ref{DNS_channels}.

Table \ref{tab:top_models} provides the model parameters
for the nine models we discuss in detail hereafter. The additional 146 models are provided in Table \ref{tab:models_add}. As discussed in what follows, we find that
the variables that have the largest impact on our results are the
He-core mass range within which a partially degenerate CO core is
formed, kick velocity dispersion, whether a HG star is allowed to
survive a CE, the value of $\alpha_{\rm CE}\lambda$, and whether hypercritical
accretion is allowed in the CE.

\subsection{DNS Evolutionary Channels}
\label{DNS_channels}

\begin{figure*}[!ht]
  \begin{center}
    \includegraphics[width=0.35\textwidth,angle=90]{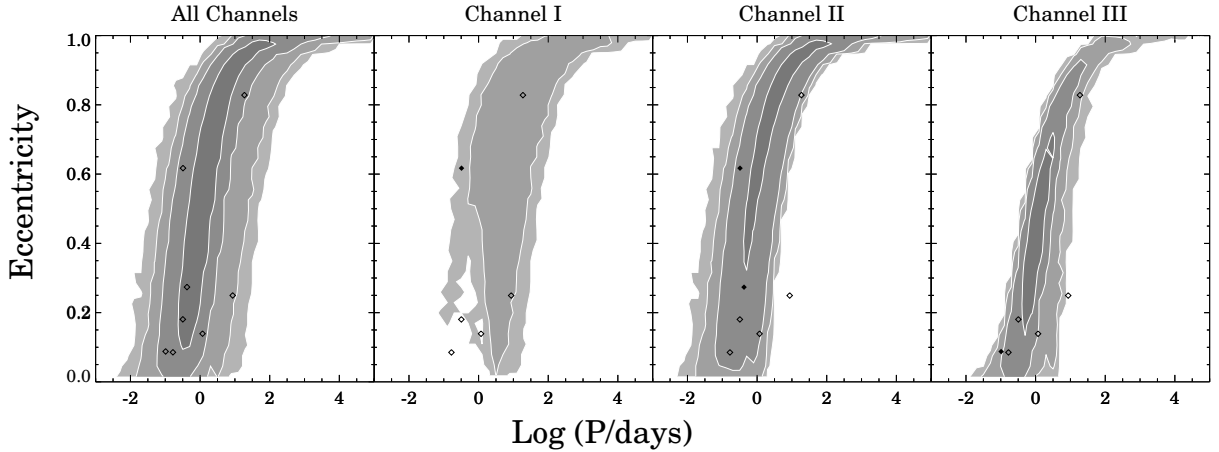}
    \caption{The DNS population of Model 1, our reference model, is
      split into evolutionary histories. The first panel includes the
      whole simulated population and the eight DNSs (open diamonds) in Table~\ref{tab:known_DNS}. The other three panels split the population into the three evolutionary channels defined in Section
      \ref{DNS_channels}. The DNSs J0737$-$3039, B1913$+$16, and B1534$+$12 are indicated by solid diamonds within their evolutionary channel restrictions described in Section \ref{evol_constrain}. 
      The four contours correspond to the 68.3\%,
      95.45\%, 99.7\%, and 99.994\% confidence
      levels, normalized to the full population. }  \label{stnd_mod}
  \end{center}
\end{figure*}

In the standard DNS formation model, the most important phases of
evolution occur after the primary becomes a NS.  Once the secondary
evolves beyond the main sequence, it may enter a phase of dynamically
unstable mass transfer.  As mentioned in \S \ref{intro}, we completely
ignore non-interacting systems because the short orbital periods
in observed DNS require mass transfer.  The secondary will evolve onto the
helium main sequence, possibly filling its Roche lobe upon evolving
off the helium main sequence in the so-called Case BB mass transfer \citep{DT81}.  After forming a degenerate core, the
star will collapse forming a NS in either an Fe core-collapse SN or
an ECS.  

Variations from the standard paradigm have been discussed in
the literature or seen in population synthesis studies
\citep{BvdH91,B95,PZY98,BKB02}. We find the scenario outlined by \citet{B95} in which the initial mass
ratio is close to unity, and the individual stars in the binary evolve
off the main sequence together, forming double common envelopes occurs
in less than 1 percent of systems, and ignore them in our analysis,
\citep[however see][]{SPR10}.  While we find some NS form through the
accretion induced collapse of a WD the population is
insignificant and we disregard them.  Taking the standard formation
paradigm outlined above, we can break down our models into three
different evolutionary tracks depending on whether or not the system went through stable mass transfer after a CE (Case BB mass transfer), and, if it did, whether the second NS was formed in an ECS or an Fe core-collapse SN.

\emph{Channel I}: As the simplest of our three formation channels, the
systems that are created this way have the fewest constraints on their
evolution.  Therefore this DNS population covers the widest range in
orbital period and eccentricity which can be seen in the second panel
in Figure \ref{stnd_mod}. Following the formation of the first NS, the
secondary evolves off the main sequence; the orbits of this population
are wide enough that the massive star fills its Roche lobe when a
significant convective envelope has developed and a CE ensues. The NS
companion is massive enough that it eventually forms an Fe core
before collapsing into the second NS in the system. 

%\emph{Channel II}: This formation scenario is similar to that of
%Channel I, but in this case, the second core collapse occurs due to an
%ECS; the secondary star has a mass within a narrow range such that
%carbon is allowed to burn, but oxygen is not. The smaller NS kick
%applied to these systems due to the secondary stars collapsing in an
%ECS, produces a population of DNS with smaller eccentricities and
%tighter orbits as can be seen in the third panel of Figures
%\ref{stnd_mod}. The progenitor mass range for ECS events is very
%narrow leading to more constrained distributions in the $P - e$ as
%shown in the third panel of Figure \ref{stnd_mod}.

\emph{Channel II}: In this channel, the binary orbit is tight enough
following the CE event that once core He burning has finished the
star expands, filling its Roche lobe again in Case BB mass transfer; 
provided the He star has not developed a fully convective envelope, mass transfer onto the first NS is stable
\citep[see ][]{IBK03}. The secondary evolves to form an Fe core and
eventually collapses in a SN.  Comparing panel two to three in Figure
\ref{stnd_mod} shows that this stable mass transfer phase produces a
DNS population with shorter orbital periods than that of Channel I.

\emph{Channel III}: These systems follow a similar evolution to Channel
II, except prior to core-collapse, the secondaries have lower He core masses, leading to NS formation through an ECS.
These lower mass He star progenitors always expand significantly leading to the second phase of mass transfer onto the primary NS. Unless the system has a very small orbital separation, this phase of mass transfer is stable \citep{DPS02,IBK03}.
Since ECS kick velocities are small, the resulting DNS orbital geometries are more dependent on the mass loss of the secondary during NS formation.
Due to the tight constraints on the mass of the secondary
immediately prior to the SN (the secondary star has a mass within a narrow range such that carbon is allowed to burn, but oxygen is not), the resulting population has a very narrow distribution of geometries.  This can be seen by the relatively narrow distribution of systems in Figure \ref{stnd_mod}.
Combined with the small kick velocities, systems going through this
evolutionary channel tend to have systematically lower
eccentricities. 

\subsection{Evolutionary Constraints}
\label{evol_constrain}

A few known DNS systems have such characteristics that a number of
different groups have been able to derive firm constraints on their
evolutionary history, and we are able to confidently place them in the
channels described here. Having this knowledge allows us to take a
step further and, for the first time, impose evolutionary constraints
on the population formation. In what follows we summarize the origin
of these evolutionary constraints on the three best studied DNS
systems. Once these constraints are imposed, we use the best models to
draw conclusions about the most likely channels through which the rest
of the DNS systems have evolved (\S \ref{indiv_systems}).

\emph{J0737$-$3039}: The binary pulsar is probably the most widely
studied DNS in terms of its evolutionary history, which is now
generally agreed upon.  After the primary star became a NS, the
secondary had to evolve off its main sequence filling its Roche lobe
in the process and forming a CE.  Both \citet{DvdH04} and
\citet{WKH04} conclude, based on the current geometry of the system,
that it must have filled its Roche lobe prior to the second SN.  Based
on the small proper motion measurement and close proximity to the
galactic disk, \citet{PS05} conclude that the He star progenitor of
J0737$-$3039B was probably $< 2.1\ M_{\odot}$, and the system received a
SN kick $\lesssim 80$ km/s at birth. This possibility, although more complex in its origin than claimed by \citet{PS05}, was confirmed by other studies \citep{WKF06,S06,vdH07,WWK10,FSK13,DPS14}.  
Limits on the He star progenitor mass for this system preclude the second born NS from having ever formed an Fe core \citep{N87}; prior to core-collapse, it was only massive enough to form an ONe core. This constraint combined with its very low natal kick requires that this system formed in an ECS.
Recently, a
population synthesis study by \citet{D10} showed that the He star
progenitor of J0737$-$3039B could not have overflowed its Roche lobe in
a second CE event; instead the mass transfer phase by the He
progenitor must have been stable.  Thus, we can conclude that
J0737$-$3039 was formed through Channel III.

\emph{B1534+12}: This system, too, has been studied quite extensively
\citep{WKH04,TDS05,S06,WWK10}. All groups agree that the only possible
way to explain all the system parameters was if the secondary had gone
through a phase of stable mass transfer as a He star.  However, this
system differs from J0737+3039 in two ways.  First, the second NS has
a mass of 1.35 $M_{\odot}$, consistent with the expected remnant of an
Fe core-collapse SN.  Second, the natal kick magnitude for this second
NS is constrained to be high, on order of several hundred km/s.  Based on our assumptions about ECS, we place B1534+12 within Channel II.

\emph{B1913+16}: Although many authors have commented on the
restrictions of the kick velocity and direction required to form
B1913+16, the evolutionary constraints on the system's formation
channel are less strong \citep{FK97,WKH04,IKB06,WWK10}.  According to
these studies, the kick velocity forming the second NS must have been
large ($\gtrsim 300$ km/s), and the mass of the secondary NS is the
expected result of an Fe core-collapse SN, so an ECS is ruled out for
the formation of the secondary NS.  However, since the mass transfer history of this system has not yet been constrained, we can only restrict
B1913+16 to have evolved through either Channel I or Channel II.

\subsection{Statistical Analysis}
\label{stats}

To quantitatively analyze the comparative goodness of fit for each
model, we adapt the Bayesian analysis used in \citet{IKB06} to fit for
two independent parameters, orbital period and eccentricity.  We begin
by applying Bayes's Theorem to our results:
\begin{equation}
{\rm P}(M|D)={{\rm P}(D|M){\rm P}(M) \over {\rm P}(D)}
\end{equation}
where ${\rm P}(M|D)$ is the probability of the model being correct
given the data, ${\rm P}(D|M)$ is the probability of the data given
the model, ${\rm P}(M)$ is the prior probability of the model, and
${\rm P}(D)$ is a normalizing constant that is independent of model,
$M$.  Our analysis, however, includes prior evolutionary constraints,
so Bayes's Theorem becomes:
\begin{equation}
{\rm P}(M|D,E) = {{\rm P}(D|M,E){\rm P}(E|M){\rm P}(M)\over {\rm P}(D,E)}
\end{equation}
where ${\rm P}(M|D,E)$ is the value which we are calculating: the
probability that our model is correct, given the data.  ${\rm
  P}(D|M,E)$ is the probability of the observed values given our
model.  ${\rm P}(E|M)$ is the probability that, given our model, the
evolutionary constraints for the data are satisfied.  ${\rm P}(M)$ is
our prior, the a priori probability that each particular model is
correct.  We give each model the same prior probability.  Let
\begin{equation}
 C={{\rm P}(M) \over {\rm P}(D,E)}
\end{equation}
yielding:
\begin{equation}
 {\rm P}(M|D,E)=C{\rm P}(D|M,E){\rm P}(E|M)
\end{equation}
Because we have given each model the same prior probability, $C$ is
independent of model.  Let $\Lambda(D)_{\rm evol}$ be 
\begin{equation}
\Lambda(D)_{\rm evol} \equiv {\rm P}(D|M,E){\rm P}(E|M).
\end{equation}
Since $C$ is independent of model, the values of $\Lambda(D)_{\rm
  evol}$ give the relative probabilities of each model.  Now, we
substitute orbital period, $P$, and eccentricity, $e$, for our data in
our equation for $\Lambda(D)_{\rm evol}$.  Due to the independence of
each observed DNS, the probability of the data set given the model is
equal to the product of the probabilities of each independent system:
\begin{equation}
\Lambda(D)_{\rm evol}=\prod_i {\rm P}(e_i,P_i|M,E_i){\rm P}(E_i|M)
\end{equation}
It is necessary to include a subscript with $E$ because there are
different evolutionary constraints on each observed system.  ${\rm
  P}(E_i|M)$ can be easily determined as the fraction of systems that
go into each evolutionary channel.

\begin{figure*}[!ht]
  \begin{center}
    \includegraphics[width=1.0\textwidth,angle=90]{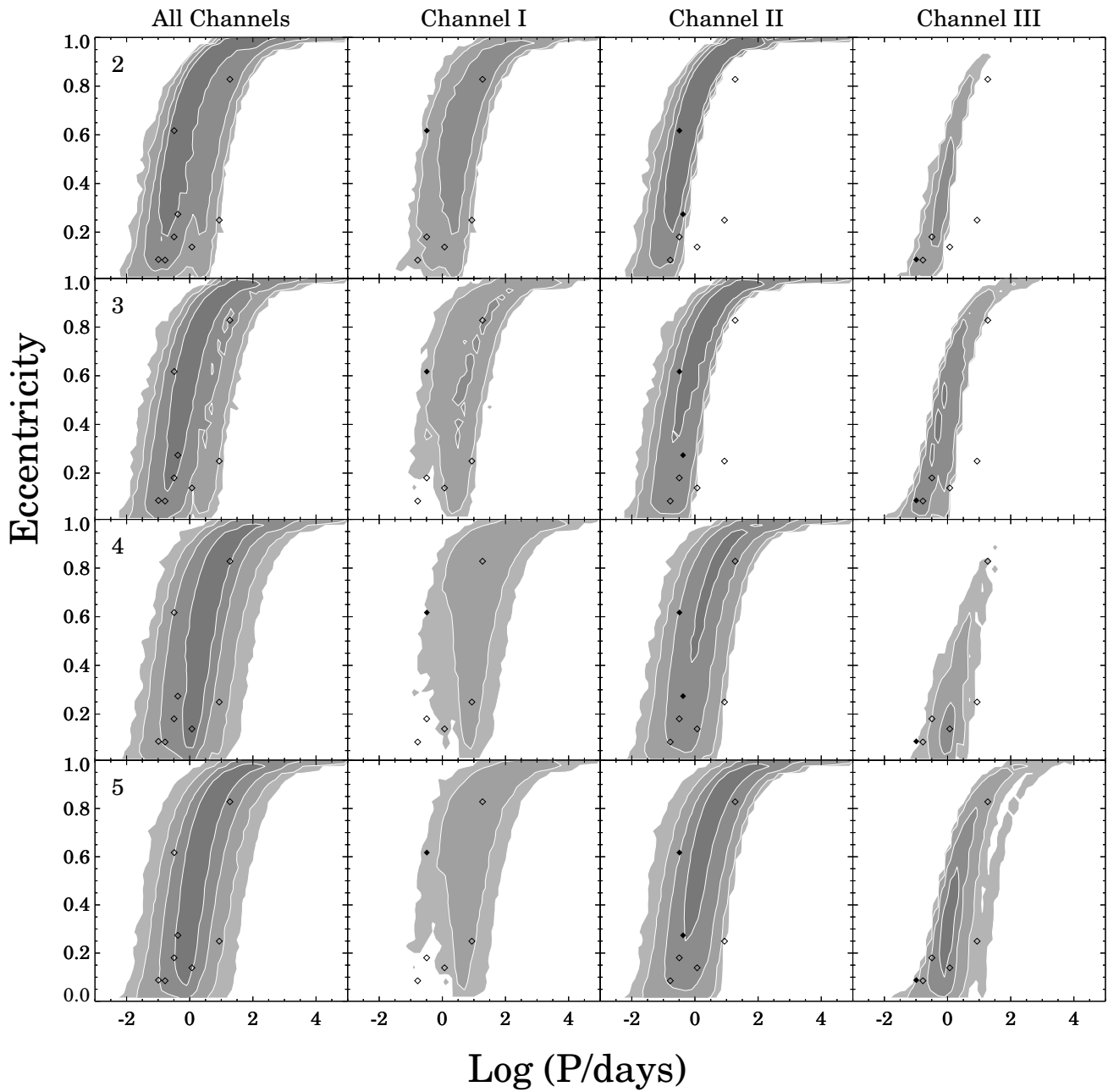}
    \caption{ The resulting population of DNSs for Models 2 through 5. Contours and symbols are the same as in Figure~\ref{stnd_mod}. Varying $\alpha_{\rm CE}\lambda$ strongly affects the distribution
      of DNS, particularly in Channel III.  The top row of
      plots shows the distribution for Model 2, with
      $\alpha_{\rm CE}\lambda$=0.25.  The second row shows the distribution for
      Model 3, with $\alpha_{\rm CE}\lambda$=0.3, while the third row shows the
      distribution for the energy conservative model, Model 4 with
      $\alpha_{\rm CE}\lambda$ = 1.0. The last row shows the distribution for Model 5, which allows NS to accrete hypercritically in a CE.}  \label{set1_fig}
  \end{center}
\end{figure*}

\begin{figure*}[!ht]
  \begin{center}
    \includegraphics[width=1.0\textwidth,angle=90]{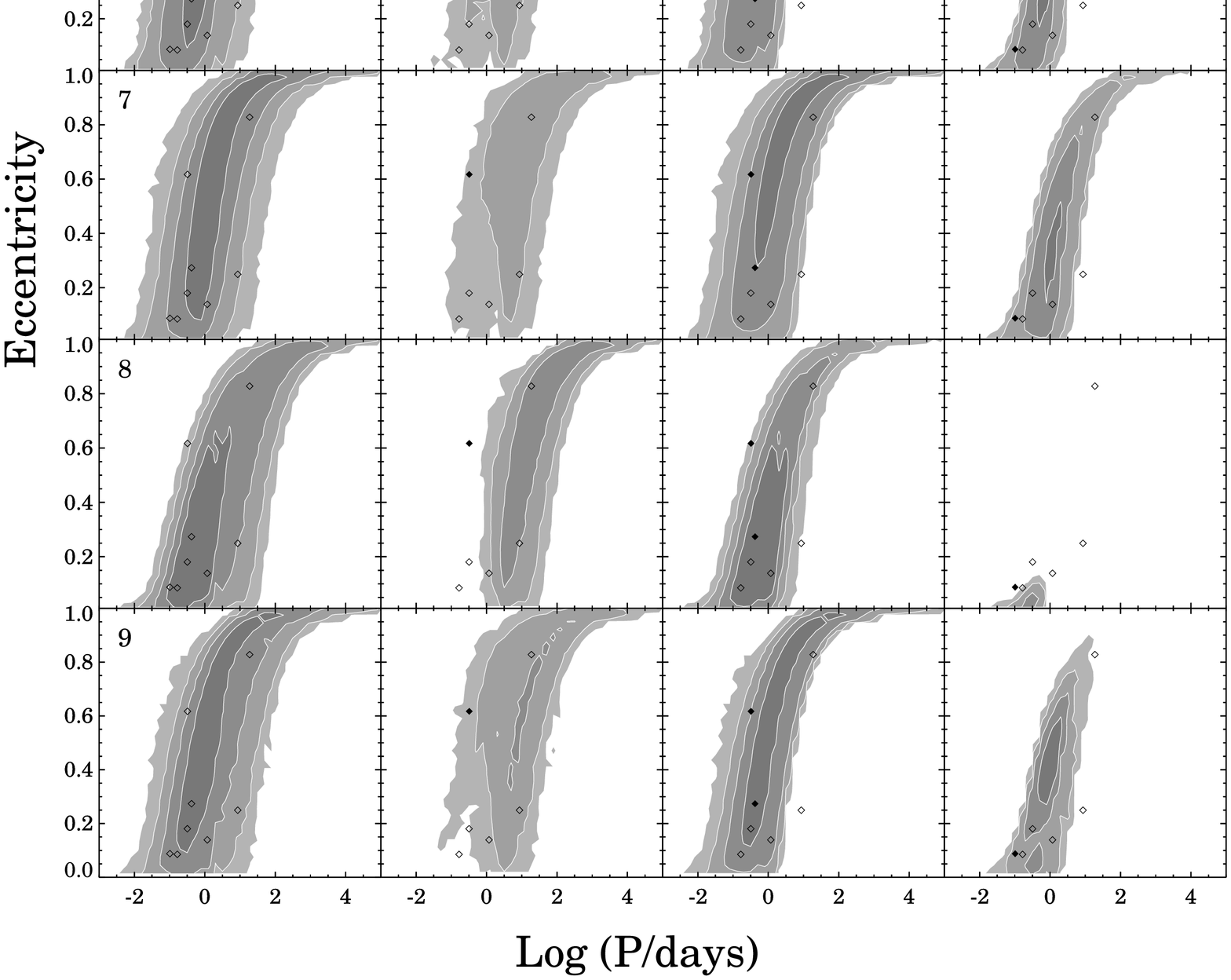}
    \caption{ The resulting population of DNSs for Models 6 through 9. Contours and symbols are the same as in Figure~\ref{stnd_mod}. Model 6 does not allow HG stars to survive a CE. Although previous studies have shown that this has a large effect on the DNS merger rate, comparison with Figure \ref{stnd_mod} shows that the qualitative difference between the two distributions is minor. Model 7 does not allow HG stars to survive a CE and also allows NS to accrete hypercritically in a CE. Models 8 and 9 show the effects of varying the Maxwellian kick velocity dispersions from 300 km/s to 50 km/s and 150 km/s, respectively.}  \label{set2_fig}
  \end{center}
\end{figure*}

The errors for the values of eccentricity and orbital period are
extremely small and can be ignored. Therefore, calculating ${\rm
  P}(e,P|M,E_i)$ reduces to finding the probability density of our
model distributions at each point in the two dimensional space
corresponding to the observed DNS. Because our data space is
two-dimensional, and each model carries no additional parameters, we
do not need to use MCMC-type methods in the computation of the model
posterior; instead we calculate ${\rm P}(e,P|M,E_i)$ by binning the
simulated systems into two dimensions for each evolutionary channel.
We use a bin size of $\Delta e = 0.033$ in eccentricity and $\Delta
\mathrm{log}(P) = 0.167$ in orbital period.  Varying the bin size gives
a good handle on the inherent error of our statistical method.  Using
equation (5), the value we are calculating, $\Lambda(D)_{\rm evol}$,
is equal to the product of the probability density of each bin that
corresponds to the observed systems.  We find that, although our
models have values of $\Lambda(D)_{\rm evol}$ that vary by several
orders of magnitude, $\Lambda(D)_{\rm evol}$ changes within a single
model by at most a factor of a few as the bin size varies over
reasonable ranges, so our model rankings are insensitive to our choice
of binning.

%The errors for the values of eccentricity and orbital period are extremely small and can be ignored. Therefore, calculating ${\rm P}(e,P|M,E_i)$ reduces to finding the probability density of our model distributions at each point in the two dimensional space corresponding to the observed DNS. We avoid using MCMC-type methods required for higher dimensional models instead favoring a simpler brute force calculation by binning the simulated systems into two dimensions for each evolutionary channel.  We use a bin size of $\Delta e = 0.01$ in eccentricity and $\Delta \mathrm{log}(P) = 0.05$ in orbital period. Varying the bin size gives a good handle on the inherent error of our statistical method.  Using equation (5), the value we are calculating, $\Lambda(D)_{\rm evol}$, is equal to the product of the probability density of each bin that corresponds to the observed systems.  We find that, although our models have values of $\Lambda(D)_{\rm evol}$ that vary by several order of magnitude, our statistical method is accurate to within a factor of a few, verifying the applicability of our method.

It is useful to determine the dependencies of our results on our use
of evolutionary constraints.  To do this, we marginalize over the nuisance parameter of the evolutionary channels. The analysis is performed similarly, the only exception being our entire model distribution is binned in two dimensions without evolutionary constraints.  In this case, $\Lambda(D)$
can be defined as:
\begin{equation}
 \Lambda(D) = \prod_i {\rm P}(e_i,P_i|M)
\end{equation}

The values for $\Lambda(D)$ (hereafter $\Lambda$) will, in general, differ from the values
we find for $\Lambda(D)_{\rm evol}$ (hereafter $\Lambda_{\rm evol}$).

\section{Results}
\label{results}

In what follows we analyze our results and draw conclusions in the
context of a number of different questions. First we examine how the
model behavior on the $P-e$ plane is affected by a few critical model
parameters. We then use the likelihood calculation for two cases, one
that includes only the $P-e$ measurements as constraints and one that
accounts for the evolutionary constraints of the three systems
discussed in \S\,3.2, and we identify the most favored models and
discuss their parameters. For these top models, we also calculate the
branching ratios through the three evolutionary channels and try to
identify the most likely channel for all eight DNS in the known
sample.

\subsection{Influence of Key Model Parameters}

The models in Figures \ref{set1_fig} and \ref{set2_fig} demonstrate
the influence of the energy efficiency in CE evolution, SN kicks, as
well as hypercritical NS accretion and survival through the CE phase
when the donor star is in the HG. 

As expected, tighter orbits are produced by models with less efficient
CE phases, those with smaller $\alpha_{\rm CE}\lambda$ values. This is reflected
in Models 2, 3, and 4 with $\alpha_{\rm CE}\lambda = 0.25$, $0.3$, and $1.0$,
respectively in Figure \ref{set1_fig}. This figure further
demonstrates the effect on branching ratios between the models;
smaller CE efficiencies lead to proportionately fewer systems becoming
DNS through Channel III. The interplay of these two effects
demonstrate the difficulty in finding models that reproduce
J0737$-$3039. 
Models 4 and 5, which also have difficulty of producing systems similar to J0737$-$3039, nonetheless remain viable models due to their increased ability to reproduce other DNSs, particularly J1518$+$4904 and J1829$+$2456. 

Smaller NS natal kick velocities result preferably in less eccentric
DNS orbits. Model 8 with a kick velocity dispersion of 50 km/s in
Figure \ref{set2_fig} shows how the highest concentrations of DNS tend
to have eccentricities smaller than 0.5. Although these low
kick models better reproduce the small eccentricity DNS, they have
difficulty reproducing B1913+16. Low kick models alter the branching
ratios, creating more DNS through Channel I. Model 9 in Figure
\ref{set2_fig}, which applies kicks with a 150 km/s dispersion, shows
an intermediate between Model 8 and the standard model.

Model 5 in Figure \ref{set1_fig} demonstrates the shift toward
slightly longer orbital periods when NS are allowed to accrete
hypercritically in a CE. Such accretion increases the NS mass, reduces
the envelope mass to be ejected and therefore requires a smaller
degree of orbital contraction for envelope ejection. On the other
hand, allowing HG stars to survive the CE phase, if there is enough
orbital energy initially, has an almost negligible effect as seen by
Model 6 in Figure \ref{set2_fig}. There is a slightly higher density
of systems at the shortest orbital periods in Channel III. Model 7,
which both allows NS to accrete hypercritically as well as allows HG
stars survive a CE, shows the combination of these two effects, as
expected.

In their study of double compact object merger rates, \citet{DBF12} find that only relatively close binaries will overfill their Roche lobe while the donor is on the HG. If these donors are allowed to survive a CE, they will produce tight binaries that merge relatively quickly due to gravitational wave radiation (as evidenced by their short calculated delay time distribution), resulting in an increased DNS merger rate. Since these systems merge soon after their formation, they are unlikely to be observed as DNS. Therefore, while this evolutionary channel may significantly contribute to the DNS merger rate, its effect on the distribution of observed DNS in $P-e$ space is negligible. 

Using a range for $M_{\rm He,deg}$ of 1.83-2.5 $M_{\odot}$ populates
Channel III, while the number of systems formed through Channel III drops precipitously for a slightly lower range of 1.83-2.25 $M_{\odot}$. We further find that while we test several different ranges, we
find that only those models with a range for $M_{\rm He,deg}$ of
1.83-2.5 $M_{\odot}$ and 2.0-2.5 $M_{\odot}$ produce DNS through
Channel III in reasonable numbers. If the He star mass range is not high enough, either the
partially degenerate CO core will not become massive enough ($>$1.08
$M_{\odot}$) to begin stable C burning, or the resulting ONe core
will not reach the Chandrasekhar limit (1.38 $M_{\odot}$) and leave an
ONe WD remnant. 

Constraints on $M_{\rm He,deg}$ were independently proposed by \citet{L09} who analyzed the population of high mass X-ray binaries in the Small Magellanic Cloud. They found that if X-ray binaries were not formed through an ECS, the numbers of observed systems with ages 20-60 Myr could not be explained. However, when using a range for $M_{\rm He,deg}$ that allowed X-ray binaries to be formed through ECS, X-ray binaries were formed with the correct age distribution. Although they did not determine the best range for $M_{\rm He,deg}$, their fiducial range of 1.83-2.25 $M_{\odot}$ compares to our preferred ranges for $M_{\rm He,deg}$ of 1.83-2.5 $M_{\odot}$ and 2.0-2.5 $M_{\odot}$. 

\subsection{Best-fit Models and their Parameters}

\begin{figure}[!ht]
  \begin{center}
    \includegraphics[width=0.9\columnwidth]{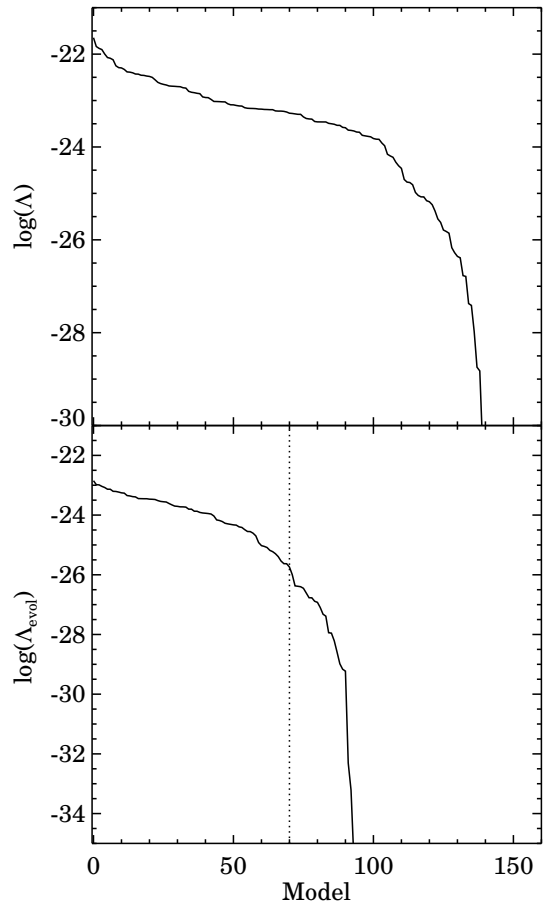}
    \caption{Our models ranked in order of decreasing likelihood. The
      top panel shows the top models for $\Lambda$, while the bottom
      panel shows the top models for $\Lambda_{\rm evol}$. The ordering
      is performed separately for each panel. The bottom panel shows
      that most of our models are not viable. We chose the top 70 for
      further analysis, shown by the dotted line in the bottom
      panel.}  \label{lamb_comb}
  \end{center}
\end{figure}

Following the analysis of \S \ref{stats}, we evaluate each of our 155
models by calculating the two likelihoods, $\Lambda$ and $\Lambda_{\rm
  evol}$, without and with the evolutionary constraints, respectively; the values
are shown in Figure \ref{lamb_comb} in decreasing order across
model number. The ordering is performed separately in each panel. 
When evolutionary constraints are included in the analysis, the number of 
viable models drops from $>$100 to 70. We select
these 70 models as their $\Lambda_{\rm evol}$ value
is within three orders of magnitude of the highest likelihood model.  Our
remaining discussion will be confined to these ``best-fit'' models indicated by
the dotted line in the bottom panel of Figure \ref{lamb_comb}. Table
\ref{tab:top_models} shows the input parameters of the top 70 models,
including the reference model.

Within the top 70 models, there is little preference either for or
against models allowing hypercritical accretion in the CE or allowing
HG stars to survive a CE. Furthermore, the secondary parameters have
little effect on the resulting distributions. These include changing
the mass beyond which a helium core is fully convective, using an
angular momentum prescription to calculate the CE evolution, altering
the efficiency of mass accretion during stable mass transfer, altering
the wind mass loss prescription and decreasing the maximum allowed
neutron star mass.

Two different ranges for $M_{\rm He,deg}$ dominate the top
models in Table \ref{tab:top_models}: 1.83-2.5 $M_{\odot}$ and 2.0-2.5
$M_{\odot}$. This is due to the difficulty for other ranges to create
DNS through Channel III, a necessity for forming J0737$-$3039. 

There is an interplay between $\alpha_{\rm CE}\lambda$ and the natal kick velocities applied in our models: low kick models push the distributions toward smaller eccentricities, while high CE efficiency models push the
distributions toward longer orbital periods. Combined, these models make it difficult to form J0737$-$3039 through Channel III. Therefore, when evolutionary constraints are applied to the models, the models most
often eliminated are those with $\alpha_{\rm CE}\lambda = 1$ and a low kick
velocity. This is consistent with independent but similar constraints derived by \citet{L09} who argue that large Fe core-collapse kicks and small ECS kicks are required to explain the population of high mass X-ray binaries within the Small Magellanic Cloud. Nevertheless, there are viable models with all three kick velocities that we test.
We note that in models with a kick dispersion velocity of 300 km/s, the bulk
of DNS are formed with $e>0.4$, while only two of the eight known DNS have such high
eccentricities. We show in \S \ref{bias} that this is unlikely to be due to a bias against detecting high eccentricity DNS.

\begin{figure}[!ht]
  \begin{center}
    \includegraphics[height=1.0\columnwidth,angle=90]{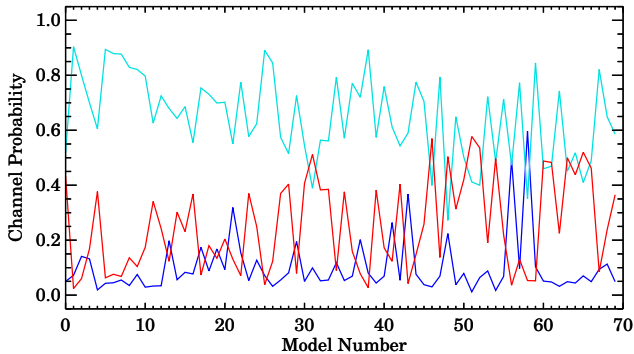}
    \caption{The branching ratios for each evolutionary channel. Models are ordered by decreasing $\Lambda_{\rm evol}$. 
Blue, teal, and red corresponds to channels I, II, and III, respectively. Channel II is the dominant formation channel,
      however many models have Channels I and III producing nearly half
      of all DNS. \label{branching}}
  \end{center}
\end{figure}

Figure \ref{branching} shows the branching ratios of the top models,
ordered by decreasing $\Lambda_{\rm evol}$. The majority of DNS in
these top models are formed through channel II shown as a green
line. These systems are characterized by partially recycled primary
NS, and secondary NS with masses of $\sim$1.35 $M_{\odot}$, consistent
with many of the DNS in Table \ref{tab:known_DNS}. The red line in
Figure \ref{branching} shows that in these models, DNS are formed in appreciable numbers through Channel III.

\subsection{Most Likely Evolutionary Channels for DNS Systems}
\label{indiv_systems}

One motivation for the present study is to examine whether we can
identify the statistically-favored evolutionary models for the DNS
systems, for which, at present, it is impossible to identify their
evolutionary history based on their individual studies. With this in
mind, we re-assess the branching ratios for the best-fit $\Lambda_{\rm evol}$ models (which have been ``calibrated'' against the systems with known evolutionary channels) 
in direct connection with the observed systems. For each observed
system and for each best-fit model, we select the sub-population of
model systems with eccentricities and orbital periods in the same 2D bin as is used to calculate our $\Lambda_{\rm evol}$ values in \S \ref{stats}. The evolutionary channel branching ratios are calculated as the proportion of systems falling within this bin that go through each evolutionary channel. Varying the bin size affects the evolutionary channel branching ratios at the 10-15\% level.

\begin{figure*}[!ht]
  \begin{center}
    \includegraphics[height=1.0\textwidth,angle=90]{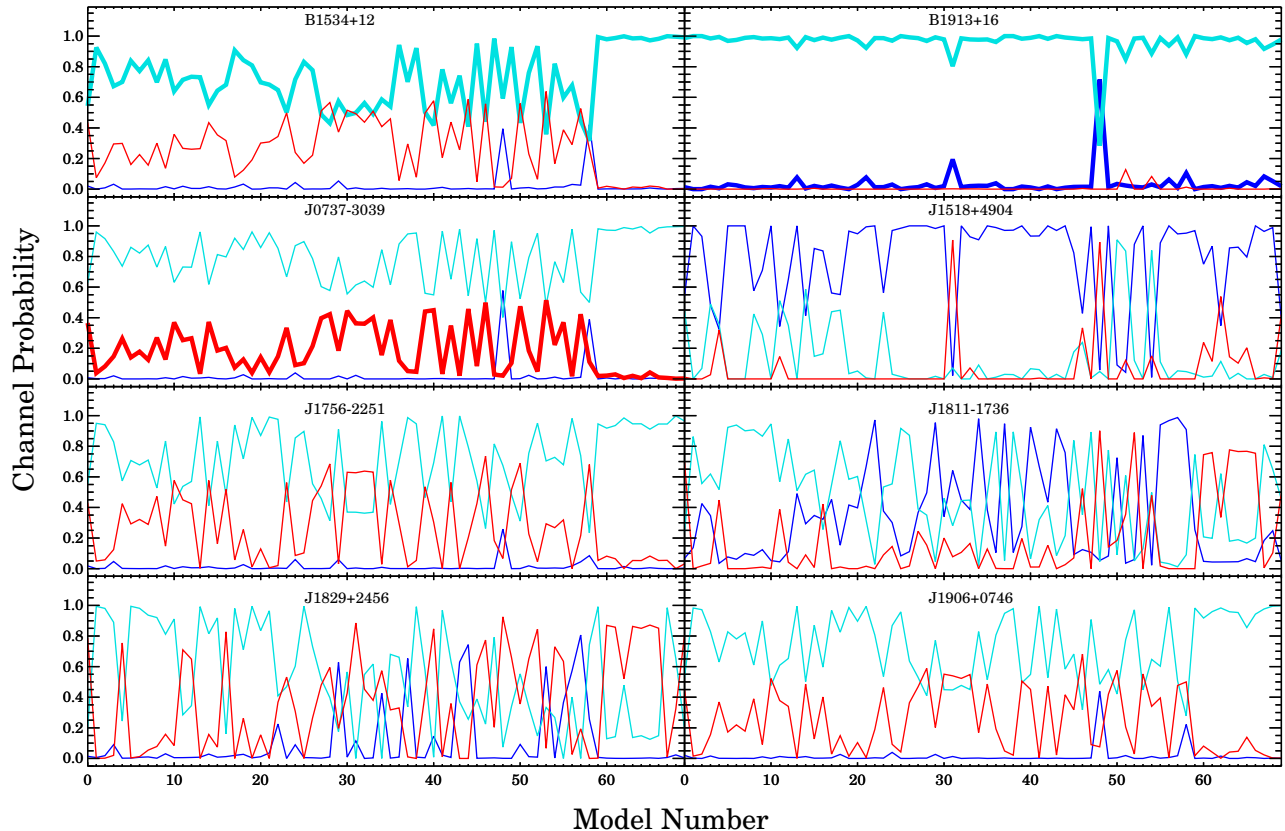}
    \caption{Blue, teal, and red correspond to channels I,
      II, and III, respectively. Again, models are ordered by decreasing $\Lambda_{\rm evol}$. Model predictions for the
      branching ratios for each DNS are shown. Our evolutionary priors
      are shown by the bold lines: teal for B1534+12, blue and teal
      for B1913+16, and red for J0737$-$3039.  \label{channels_top}}
  \end{center}
\end{figure*}

The results for all 70 viable models are shown in Figure \ref{channels_top}. For systems such as B1913+16 and J1518+4904, one channel is clearly favored {\em statistically}, although this is not necessary the channel that actually formed each system. A few of these models are peculiar. The models at 31 and 48 are models that calculate CE evolution based on the conservation of angular momentum not energy. Although our statistical analysis gives each model an equal prior, modern discussions generally disfavor this prescription \citep{W08,WIv11,IJC13}.

\emph{J0737$-$3039:} This is the prime example illustrating that the
\emph{statistically favored} channel is not necessarily the
\emph{true} channel. In this case, the former is Channel II, while
the latter is Channel III, based on the analysis of all its observed
characteristics (as discussed in \S\,3.2). In Figure
\ref{channels_top}, the red line representing Channel III shows that in all models, no more than half of all DNS are formed through this pathway. 
%Nevertheless, the DNS that are formed through this channel tend to have small eccentricities and orbital periods, as was seen in Figures \ref{set1_fig} and \ref{set2_fig}. This is reflected in Figure \ref{channels_top}, where although none of the models here have a branching ratio larger than $\sim$25\% for Channel IV for J0737-3039, the branching ratio is much higher than that seen in Figure \ref{branching}.

\emph{B1534+12:} The requirement of B1534+12 to form through Channel
II is in agreement with the branching ratio results shown in Figure
\ref{channels_top}; in this case the \emph{statistically favored}
channel is also the \emph{true} channel. As discussed in \S\,3.2 the
mass measurements and NS kick constraints indicate that B1534+12 could
not have been formed in an ECS, precluding formation through Channel
III, the other likely evolutionary channel in Figure
\ref{channels_top}.

\emph{B1913+16:} Prior evolutionary constraints restrict B1913+16 to
be created through either Channel I or II. In Figure
\ref{channels_top} we see that, of these two, Channel II is greatly
favored from a statistical point of view. The models at 31 and 48 correspond to models that calculate CE evolution based on an angular momentum conservation description. Hence we can say that
B1913+16 \emph{most likely} went through a phase of stable mass
transfer prior to the second SN event. This can be attributed to both
Channel II having a much higher branching ratio as well as the
systems going through Channel I being formed at larger orbital periods.

\emph{J1756$-$2251:} Recent observations by \citet{FSK14} indicate the pulsar companion in J1756$-$2251 has a mass of 1.23 $M_{\odot}$. The low mass combined with their newly measured tangential velocity of $\sim$20 km s$^{-1}$ suggests that the pulsar companion may have been
born in an ECS. If that is the case, J1756$-$2251 must have been formed
through Channel III. The branching ratio for Channel III
(red in Figure \ref{channels_top}) varies between 10\% and 60\% for
J1756$-$2251, indicating this possibility. If we add the constraint requiring J1756$-$2251 to be formed through Channel III, the resulting $\Lambda_{\rm evol}$ values do not change significantly.

\emph{J1906+0746:} J1906+0746 is unique in that the observed pulsar is
the second-born NS in the system. A mass measurement for the pulsar
indicates it may have been born in an ECS, constraining the system to
Channel III. Figure \ref{channels_top} indicates that this system almost certainly went through either Channel II or III. Through future analysis of the galactic position and velocity of J1906+0746, it may be possible to determine if it went
through a phase of stable mass transfer, and therefore Channels II or III. However, with the available information we cannot differentiate between the two channels, as the branching ratios in Figure \ref{channels_top} indicate both are possible.

\emph{J1811$-$1736:} The branching ratios in Figure \ref{channels_top} indicate that J1811$-$1736 likely evolved through Channel I or II. Its relatively large orbital period means formation through Channel III is rare for most models. Models 7, 8, and 9 in Figure \ref{set2_fig} show this is because of the difficulty of forming a DNS with large orbital periods and eccentricities through the low velocity kick applied in an ECS. The models that allow for the formation of J1811$-$1736 through Channel III tend to be those with higher kick velocities, and hence larger ECS kicks. Based on their measurements of the relativistic periastron advance and the low derived system mass, \citet{CKS07} suggest that J1811$-$1736 was born with a low velocity kick. Although uncertainties in the current mass measurements for the system are too large to determine evolutionary constraints, \citet{CKS07} calculate that a second post-Keplerian parameter may be measureable in the near future. If future observations indicate this system was formed through Channel III, it could provide an important constraint on ECS kicks.

\emph{J1518+4904:} The $\sim$8.6 day orbital period and eccentricity
of 0.249 for J1518+4904 make its formation difficult. The
distributions in Figures \ref{set1_fig} and \ref{set2_fig} show that most DNS
are formed at either smaller orbital periods or larger
eccentricities. Although a few models form J1518+4904
through Channel II or III, Figure \ref{channels_top} indicates it was most likely formed through Channel I. Observational errors on the mass estimates are too large to indicate whether the companion was born in an ECS event or not.

\emph{J1829+2456:} Depending on the individual model, Figure
\ref{channels_top} shows that J1829+2456 formed through any of
the three evolutionary channels. Poor mass constraints do not allow any
further indication. Future observations may provide further
constraints on its evolution.

\section{Discussion}
\label{discussion}

\begin{figure*}[!ht]
  \begin{center}
    \includegraphics[width=0.9\textwidth]{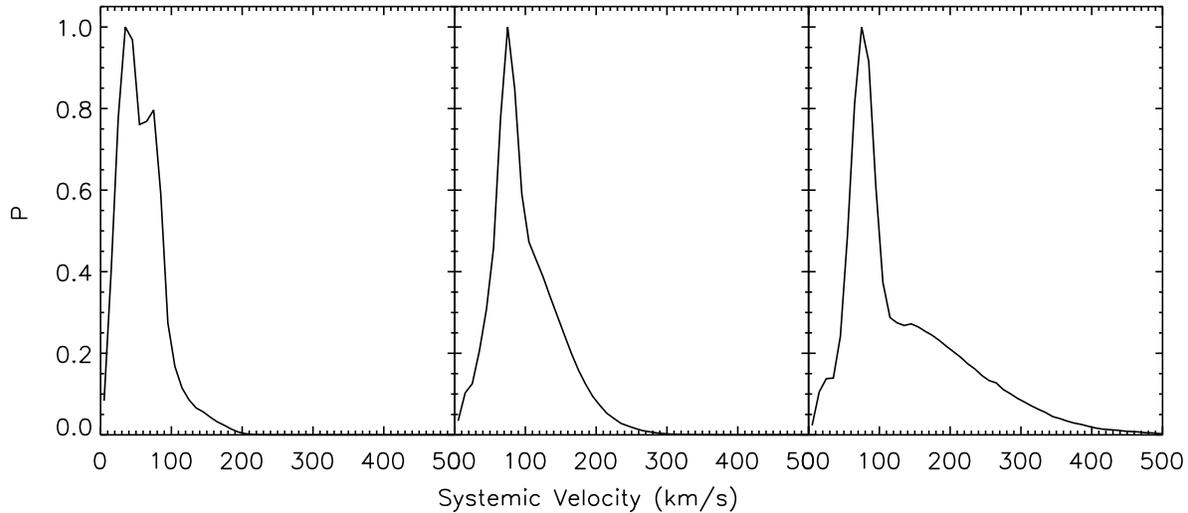}
    \caption{ The distribution of systemic velocities of the DNS formed by (from left to right) Model 8, 9, and 1. Model 1 with a Maxwellian kick dispersion velocity of 300 km/s produces a tail of DNS with systemic velocities extending up to 400 km/s. However, the bulk of the DNS are formed with systemic velocities less than 200 km/s. \label{sys_vel_hist}}
  \end{center}
\end{figure*}

\begin{figure}[!ht]
  \begin{center}
    \includegraphics[height=0.75\columnwidth]{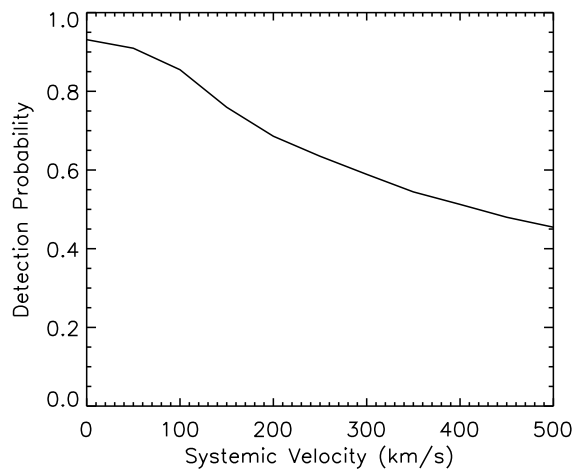}
    \caption{ The probability at which a DNS will fall within the Parkes multibeam pulsar survey region ($|b| < 5^{\circ}$ and $260^{\circ} < l < 50^{\circ}$) as a function of the velocity applied to the system. The details of our Galactic model and our calculation of the detection probability are provided in \S \ref{bias}. \label{sys_vel}}
  \end{center}
\end{figure}

\subsection{Partial Recycling}

While there is little observational evidence to distinguish between
different evolutionary channels, one such piece of evidence could be
the degree to which the first-born NS has been recycled. \citet{I11}
has argued that the mass losing star in a CE will go
through a phase of thermal readjustment upon envelope
expulsion. During this phase a NS companion can be mildly
recycled. Investigation of Table 1 shows that there may be in fact a
bifurcation spin periods in the primary NS: those with $P_s \lesssim$
50 ms and those with $P_s \gtrsim$ 100 ms. If this argument is
accurate, then J1811$-$1736 went through Channel I,
avoiding any stable mass transfer after the CE phase, while the other
systems with shorter spin periods went through Channel II or
III. J1906+0746 is an exception because we do not know the spin period
of the primary NS.

Comparison with Figure \ref{channels_top} shows that this idea is
consistent with most of the results of our analysis here. In roughly
half of our viable models, Channel I is preferred for J1811$-$1736. With a few exceptions, other systems prefer Channel II (in agreement with their shorter spin
periods) except for J1518+4904 which our results here indicate was
likely formed through Channel I. Either this system, too, formed
through Channel II, or the primary NS in J1518+4904 accreted enough
material in a CE to recycle the primary to its current spin
period of 40.9 ms.

\subsection{Observational Biases}
\label{bias}

Our statistical analysis discussed in \S\ref{stats} relies on the
assumption that all DNS have an equal detection likelihood. For
example, a strong bias against detection of DNS with large
eccentricities means that high kick velocity models, which produce on average more eccentric DNS, will not be as disfavored as the $\Lambda_{\rm evol}$ values indicate. In principle our DNS distributions could be convolved with any observational bias for a more accurate comparison to the observed DNS.

One such bias deals with the observability of pulsars in DNS which is a function of a number of intrinsic parameters such as the beaming fraction, magnetic field strength, age, and spin period as well as particular parameters such as the distance and direction from the Sun. However, since we ignore non-interacting DNS, our model assumes that every DNS contains a partially recycled pulsar, each with an equal probability of being observed. Absent any deeper understanding of the connection between a pulsar's characteristics and its prior evolution, we take this to be a reasonable approximation.

An additional bias could be caused by Doppler smearing of pulsars in short
orbital periods.  Due to the high accelerations in such systems over
the course of a single observation, close binaries suffer a decreased
detection efficiency. \citet{BLW13} analyzed the detectability of binary pulsars in surveys, finding that modern acceleration searches (e.g. PRESTO) were greater than 80\% efficient at detecting pulsars in DNS for typical parameters. Interestingly, they found systems with the shortest orbital periods and lowest eccentricities suffered the strongest biases, which still have a detection efficiency greater than 50\%. 

A potentially more pernicious bias could be introduced by the limited sky coverage of pulsar surveys, typically close to the Galactic plane. Binaries that gain large systemic velocities due to the NS natal kicks may therefore be missed.  Such DNS would likely have significant
eccentricity and a large orbital period, possibly helping to explain
the relative dearth of such systems despite the predictions by our
simulations. We first investigate the systemic velocities of our resulting DNS populations.

The distribution of systemic velocities, normalized to the highest bin, for Models 8 ($\sigma_{\rm kick}=50$ km/s), 9 ($\sigma_{\rm kick}=150$ km/s), and 1 ($\sigma_{\rm kick}=300$ km/s) are shown in Figure \ref{sys_vel_hist}. We expect these three models to be representative of the different kick velocity models of our entire set. Calculated from the equations in \citet{K96}, this systemic velocity arises from both the mass lost from the collapsing object as well as the kick applied to the newly born NS. Model 8 produces DNS with systemic velocities peaking at 35 km/s. Model 9 and 1 both produce DNS with systemic velocities peaking at 75 km/s, although the higher kick velocity Model 1 has a high velocity tail. Can these systemic velocities create a bias against detection strong enough to significantly affect our quantitative results?

\begin{figure*}[!ht]
  \begin{center}
    \includegraphics[height=0.65\textwidth, angle=90]{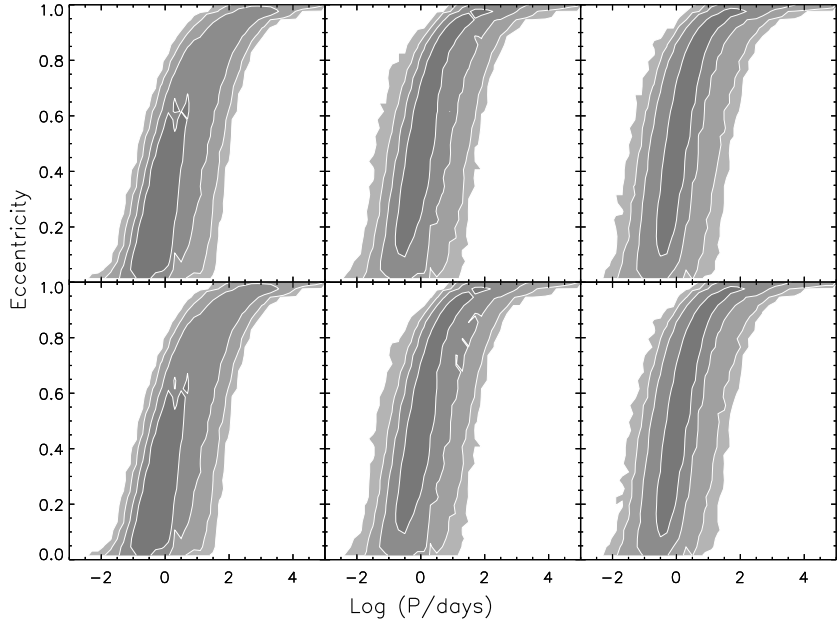}
    \caption{ The top panels show the distributions of DNS in $P-e$ space for models 8, 9, and 1, with Maxwellian kick dispersion velocities of 50, 150, and 300 km/s, respectively. The bottom panels show the distributions from the same models, convolved with the detection probability shown in Figure \ref{sys_vel}. It is evident that the differences are minimal. \label{bias_convolved}}
  \end{center}
\end{figure*}

To determine this magnitude of any effect, we create a model Milky Way-like galaxy composed of a three-component potential defined in Appendix A of \citet{WKK00}. DNS are born in a double exponential, axisymmetric disk with a scale length of 2.80 kpc and scale height of 70 pc, and are given an initial velocity corresponding to circular rotation. The motion of the DNS through the Galaxy is calculated using a Runge-Kutta fourth-order scheme. The binary is evolved for 20 Myr (the approximate evolution time of the binary prior to the birth of the second NS) before a systemic velocity is applied. The systemic velocity is applied to the system in a randomly chosen direction. The system is then evolved for a Gyr and the location of the binary recorded throughout this evolution at intervals of a Myr. This location is translated into a galactic longitude and latitude as referenced from the Sun, placed at 8.5 kpc from the center of the Galaxy. We repeat this process, simulating the motion of 1000 binaries throughout the Galaxy, each with a different randomly-chosen initial position and systemic velocity direction. We then determine the percentage of the resultant Galactic longitudes and latitudes that fall within a region near the Galactic Plane, using as a fiducial region the Parkes multibeam pulsar survey area \citep{M01}: $|b| < 5^{\circ}$ and $260^{\circ} < l < 50^{\circ}$. Figure \ref{sys_vel} shows this percentage over the relevant range of systemic velocities. We find that when no velocity is applied, a DNS has a 93\% chance of falling within the Parkes multibeam pulsar survey region. That number falls to 85\% for systemic velocities of 100 km/s and 69\% for systemic velocities of 200 km/s. 

We test the strength of this bias on Models 8, 9, and 1 with Maxwellian kick dispersion velocities of 50, 150, and 300 km/s, respectively. For each simulated system in the three models, we find the detection probability from Figure \ref{sys_vel} corresponding to its individual systemic velocity. We then convolve observational biases into our distributions in $P-e$ space by selecting a subset of our original distribution. Each simulated DNS has a chance equal to its calculated detection probability of being included into the subset. In Model 1 (with the highest kick velocities and hence the strongest expected effects of the three models) roughly 80\% of all systems are unaffected. The effects of this observational bias are seen in Figure \ref{bias_convolved}; the top panels provide the full distributions of DNS for our three test models while the bottom panels show the distributions convoluted with the observational bias. The resulting distributions are nearly qualitatively identical. Since our quantitative results are robust to a factor of several orders of magnitude, we can safely ignore this observational bias as potentially altering our conclusions in this study.

\section{Summary}
\label{sum}

We generate populations of DNS using a binary population synthesis
code. In total, we test 155 models. The simulated distributions are
broadly consistent with the orbital periods and eccentricities of the
known systems. These distributions are split into three separate
evolutionary channels based on two criteria: if the secondary star
entered a phase of stable Roche lobe overflow as a He star and if the
secondary formed a NS in an Fe core-collapse SN or an ECS. Previous
authors have constrained the evolutionary histories of three of the
eight known DNS: B1913+16, B1534+12, and J0737$-$3039.

For the first time, we combine evolutionary constraints on the three
known systems with a Bayesian analysis to quantitatively estimate the
relative likelihood of each model. Of our 155 total models, we find
that 70 can create B1913+16, B1534+12, and J0737$-$3039 through
their constrained evolutionary channels.

We use these 70 viable models to further constrain binary evolution
parameters. 
We find that a common envelope efficiency $\alpha_{\rm CE}\lambda \lesssim 0.25$ is effectively ruled out. 
Although values between 0.3
and 1.0 are allowed, 0.5 tends to fit the data best. 
Since evolutionary codes typically find $\lambda<1$ for the high-mass AGB donor stars within a CE, DNS formation may require efficient CE evolution. More work is required to determine if this requires $\alpha_{\rm CE}$ greater than unity. 
The most
constraining parameter is $M_{\rm He,deg}$, the He core mass which
allows the formation of a partially degenerate CO core. Of all the
mass ranges we test, only 1.83-2.5 $M_{\odot}$ and 2.0-2.5 $M_{\odot}$
allow for the formation of J0737$-$3039 in reasonable numbers. We test models with NS natal kicks drawn from a Maxwellian distribution. Models survive with all three tested dispersion velocities (50, 150, and 300 km/s). 

We can further use our 70 viable models to constrain which
evolutionary channels formed each of the eight known DNS. We find that
B1534+12 likely went through Channel II, in agreement with its
evolutionary constraint. Although previous work constrains B1913+16 to have gone
through either Channel I or II, our results here indicate it most likely went through Case BB mass transfer in our Channel
II. Although difficult to form in general, J0737-3039 could have been
formed through Channel III, in agreement with its evolutionary
constraint. J1756$-$2251 and J1906+0746 may have been formed through Channel
III, consistent with the idea that ECS produce low mass NS. J1811$-$1736 likely went through Channel I or II, however if future observations and analysis indicate it was born through Channel III, it could place a strong constraint on ECS kick velocities. Our models indicate J1518+4904 most likely formed through Channel I. It is clear that understanding the evolutionary history of known DNS systems provides strong constraints on population models and helps us further constrain NS formation. 

As already discussed J0737$-$3039 appears to have formed through channel III, which systematically has a lower branching ratio compared to channel II (no more than about 50\% as shown in Figure \ref{channels_top}). These low branching ratio values may seem in contrast with early indications that the formation rates of pulsar binaries similar to J0737$-$3039 dominate the total DNS by a factor of 6-7 \citep{KKL04}. However, the most recent analysis of the double pulsar system \citep{KBM13} concludes that current beaming constraints lead to the empirical DNS rates being comparable for J0737$-$3039 and B1913+16. 

\acknowledgments

We would like to thank the anonymous referee for greatly improving our manuscript. We would also like to thank Paul Kiel, Tim Linden, Mia Ihm, and Ilya Mandel
for useful discussions. WF acknowledges support from a CIERA Postdoctoral Fellowship. VK acknowledges support through a Simons Foundation Fellowship in Theoretical Astrophysics and also is thankful for the hospitality provided by the Aspen Center for Physics while working on this project. 

\def\nar{{New A}}
\bibliographystyle{apj}
\bibliography{references}

%\begin{thebibliography}

%\end{thebibliography}

\vspace{1in}
%\clearpage

\input{table3.tex}

\end{document}

%% file: table3.tex
\LongTables

\begin{deluxetable*}{clllllllll}
\tablecaption{All Models \label{tab:models_add}}
\tablehead{
\colhead{Model} & 
\colhead{ECS Range\tablenotemark{a}} & 
\colhead{$\alpha_{\rm CE}$} & 
\colhead{Kick\tablenotemark{b}} & 
\colhead{HCE\tablenotemark{c}} & 
\colhead{CE\tablenotemark{d}} & 
\colhead{Notes\tablenotemark{e}} & 
\colhead{log($\Lambda$)} & 
\colhead{log($\Lambda_{\rm evol}$)} & 
\colhead{Rank\tablenotemark{f}} \\ 
 & \colhead{[$M_{\odot}$]} & & 
\colhead{[km/s]} & & & & & &
}
\startdata
10 & 1.83-2.25 & 0.3 & 150 & on & off &  & -23.1 & -50.0 & 140 \\ 
11 & 1.83-2.25 & 0.3 & 300 & on & off &  & -23.7 & -50.0 & 139 \\ 
12 & 1.83-2.25 & 0.3 & 50 & on & off &  & -23.2 & -50.0 & 138 \\ 
13 & 1.83-2.25 & 0.5 & 150 & on & off &  & -23.2 & -50.0 & 137 \\ 
14 & 1.83-2.25 & 0.5 & 300 & on & off &  & -23.6 & -50.0 & 136 \\ 
15 & 1.83-2.25 & 0.5 & 50 & on & off &  & -23.8 & -50.0 & 135 \\ 
16 & 1.83-2.25 & 1.0 & 150 & on & off &  & -23.2 & -50.0 & 134 \\ 
17 & 1.83-2.25 & 1.0 & 300 & on & off &  & -23.5 & -50.0 & 133 \\ 
18 & 1.83-2.25 & 1.0 & 50 & on & off &  & -23.5 & -50.0 & 132 \\ 
19 & 1.83-2.25 & 0.3 & 150 & off & off &  & -23.2 & -50.0 & 131 \\ 
20 & 1.83-2.25 & 0.3 & 300 & off & off &  & -24.7 & -26.8 & 79 \\ 
21 & 1.83-2.25 & 0.3 & 50 & off & off &  & -21.9 & -50.0 & 130 \\ 
22 & 1.83-2.25 & 0.5 & 150 & off & off &  & -22.3 & -50.0 & 129 \\ 
23 & 1.83-2.25 & 0.5 & 300 & off & off &  & -23.2 & -50.0 & 128 \\ 
24 & 1.83-2.25 & 0.5 & 50 & off & off &  & -21.8 & -50.0 & 127 \\ 
25 & 1.83-2.25 & 1.0 & 150 & off & off &  & -22.6 & -50.0 & 126 \\ 
26 & 1.83-2.25 & 1.0 & 300 & off & off &  & -23.0 & -50.0 & 125 \\ 
27 & 1.83-2.25 & 1.0 & 50 & off & off &  & -22.7 & -50.0 & 124 \\ 
28 & 1.83-2.5 & 1.0 & 300 & on & off &  & -23.5 & -50.0 & 123 \\ 
{\bf 29} & 1.83-2.5 & 0.3 & 150 & on & off &  & -22.5 & -23.1 & 7 \\ 
{\bf 30} & 1.83-2.5 & 0.3 & 300 & on & off &  & -22.9 & -23.3 & 13 \\ 
31 & 1.83-2.5 & 0.3 & 50 & on & off &  & -22.7 & -23.6 & 28 \\ 
32 & 1.83-2.5 & 0.5 & 150 & on & off &  & -23.2 & -23.9 & 42 \\ 
33 & 1.83-2.5 & 0.5 & 300 & on & off &  & -23.6 & -24.3 & 52 \\ 
34 & 1.83-2.5 & 0.5 & 50 & on & off &  & -23.8 & -25.5 & 68 \\ 
35 & 1.83-2.5 & 1.0 & 150 & on & off &  & -23.4 & -50.0 & 122 \\ 
36 & 1.83-2.5 & 1.0 & 50 & on & off &  & -24.3 & -50.0 & 121 \\ 
37 & 1.83-2.5 & 1.0 & 300 & off & off &  & -23.5 & -25.0 & 62 \\ 
38 & 1.83-2.5 & 0.2 & 300 & off & off &  & -25.2 & -26.5 & 76 \\ 
39 & 1.83-2.5 & 0.3 & 300 & off & off &  & -23.1 & -23.9 & 41 \\ 
40 & 1.83-2.5 & 0.3 & 150 & off & off &  & -22.8 & -23.8 & 36 \\ 
41 & 1.83-2.5 & 0.3 & 50 & off & off &  & -22.7 & -23.9 & 40 \\ 
42 & 1.83-2.5 & 0.3 & 300 & off & off & F$_{\rm a}$ = 1 & -23.5 & -24.7 & 59 \\ 
43 & 1.83-2.5 & 0.3 & 300 & off & off & $M_{\rm He,con}$ = 2 & -23.3 & -24.2 & 47 \\ 
44 & 1.83-2.5 & 0.3 & 300 & off & off & Wind 1 & -24.0 & -24.5 & 56 \\ 
{\bf 45} & 1.83-2.5 & 0.5 & 150 & off & off &  & -22.4 & -23.4 & 18 \\ 
46 & 1.83-2.5 & 0.5 & 300 & off & off &  & -23.4 & -23.9 & 39 \\ 
47 & 1.83-2.5 & 0.5 & 50 & off & off &  & -22.1 & -23.5 & 25 \\ 
48 & 1.83-2.5 & 1.0 & 150 & off & off &  & -23.2 & -50.0 & 120 \\ 
49 & 1.83-2.5 & 1.0 & 50 & off & off &  & -23.1 & -50.0 & 119 \\ 
50 & 1.83-2.5 & 1.0 & 300 & off & off & F$_{\rm a}$ = 1 & -23.3 & -50.0 & 118 \\ 
51 & 1.83-2.5 & 1.0 & 300 & off & off & $M_{\rm He,con}$ = 2 & -23.5 & -25.4 & 67 \\ 
52 & 1.83-2.5 & 1.0 & 300 & off & off & Wind 1 & -24.2 & -25.3 & 66 \\ 
53 & 1.83-2.5 & 1.0 & 300 & off & off & $\beta$ = 3 & -23.7 & -25.0 & 61 \\ 
54 & 1.83-2.5 & 1.0 & 300 & off & off & F$_{\rm a}$ = 0 & -26.2 & -26.9 & 81 \\ 
55 & 1.83-2.5 & 1.0 & 300 & off & off & $M_{\rm He,con}$ = 3.5 & -23.6 & -25.2 & 65 \\ 
56 & 2.0-2.5 & 1.0 & 300 & on & off &  & -23.3 & -50.0 & 117 \\ 
{\bf 57} & 2.0-2.5 & 0.3 & 150 & on & off &  & -22.4 & -23.2 & 10 \\ 
{\bf 58} & 2.0-2.5 & 0.3 & 300 & on & off &  & -22.7 & -23.2 & 9 \\ 
59 & 2.0-2.5 & 0.3 & 50 & on & off &  & -22.7 & -23.9 & 38 \\ 
60 & 2.0-2.5 & 0.5 & 150 & on & off &  & -23.0 & -24.3 & 51 \\ 
61 & 2.0-2.5 & 0.5 & 50 & on & off &  & -23.6 & -24.9 & 60 \\ 
62 & 2.0-2.5 & 1.0 & 150 & on & off &  & -23.0 & -50.0 & 116 \\ 
63 & 2.0-2.5 & 1.0 & 50 & on & off &  & -23.7 & -50.0 & 115 \\ 
{\bf 64} & 2.0-2.5 & 1.0 & 300 & on & on &  & -22.7 & -23.3 & 12 \\ 
65 & 2.0-2.5 & 0.3 & 300 & on & on &  & -22.9 & -23.6 & 27 \\ 
{\bf 66} & 2.0-2.5 & 0.3 & 150 & on & on &  & -22.3 & -23.1 & 6 \\ 
{\bf 67} & 2.0-2.5 & 0.3 & 50 & on & on &  & -21.6 & -23.0 & 4 \\ 
{\bf 68} & 2.0-2.5 & 0.5 & 150 & on & on &  & -22.4 & -23.2 & 8 \\ 
{\bf 69} & 2.0-2.5 & 0.5 & 50 & on & on &  & -22.1 & -23.0 & 3 \\ 
70 & 2.0-2.5 & 1.0 & 150 & on & on &  & -22.6 & -23.5 & 24 \\ 
{\bf 71} & 2.0-2.5 & 1.0 & 50 & on & on &  & -22.5 & -23.4 & 17 \\ 
{\bf 72} & 2.0-2.5 & 1.0 & 300 & off & on &  & -22.3 & -22.8 & 1 \\ 
73 & 2.0-2.5 & 0.1 & 300 & off & on &  & -50.0 & -50.0 & 114 \\ 
74 & 2.0-2.5 & 0.2 & 300 & off & on &  & -50.0 & -50.0 & 113 \\ 
75 & 2.0-2.5 & 0.3 & 300 & off & on &  & -23.7 & -24.6 & 58 \\ 
76 & 2.0-2.5 & 0.3 & 150 & off & on &  & -22.7 & -23.7 & 34 \\ 
77 & 2.0-2.5 & 0.3 & 50 & off & on &  & -22.7 & -24.6 & 57 \\ 
{\bf 78} & 2.0-2.5 & 0.5 & 150 & off & on &  & -22.0 & -23.0 & 2 \\ 
79 & 2.0-2.5 & 0.5 & 50 & off & on &  & -22.1 & -23.5 & 23 \\ 
80 & 2.0-2.5 & 1.0 & 150 & off & on &  & -22.4 & -23.5 & 22 \\ 
81 & 2.0-2.5 & 1.0 & 50 & off & on &  & -22.6 & -23.9 & 37 \\ 
{\bf 82} & 2.0-2.5 & 1.0 & 300 & on & on & $M_{\rm NS,max}$ & -22.5 & -23.1 & 5 \\ 
83 & 2.0-2.5 & 0.1 & 300 & off & off &  & -27.4 & -50.0 & 112 \\ 
84 & 2.0-2.5 & 0.2 & 300 & off & off &  & -25.2 & -26.4 & 74 \\ 
85 & 2.0-2.5 & 0.3 & 150 & off & off &  & -22.5 & -23.5 & 21 \\ 
86 & 2.0-2.5 & 0.3 & 50 & off & off &  & -22.3 & -24.0 & 44 \\ 
87 & 2.0-2.5 & 0.5 & 300 & off & off & $\beta$ = 3 & -22.9 & -23.6 & 26 \\ 
88 & 2.0-2.5 & 0.5 & 300 & off & off & F$_{\rm a}$ = 0.1 & -23.9 & -24.5 & 55 \\ 
89 & 2.0-2.5 & 0.5 & 300 & off & off & F$_{\rm a}$ = 0 & -23.5 & -24.3 & 50 \\ 
90 & 2.0-2.5 & 0.5 & 300 & off & off & F$_{\rm a}$ = 0.3 & -23.8 & -24.2 & 45 \\ 
91 & 2.0-2.5 & 0.5 & 300 & off & off & F$_{\rm a}$ = 1 & -23.8 & -50.0 & 111 \\ 
{\bf 92} & 2.0-2.5 & 0.5 & 300 & off & off & $M_{\rm NS,max}$ & -22.8 & -23.5 & 19 \\ 
93 & 2.0-2.5 & 0.5 & 300 & off & off & $M_{\rm He,con}$ = 2.5 & -23.2 & -23.7 & 32 \\ 
94 & 2.0-2.5 & 0.5 & 300 & off & off & $M_{\rm He,con}$ = 2 & -23.2 & -23.7 & 33 \\ 
95 & 2.0-2.5 & 0.5 & 300 & off & off & $M_{\rm He,con}$ = 3.5 & -23.2 & -23.7 & 31 \\ 
96 & 2.0-2.5 & 0.5 & 300 & off & off & $\gamma_{\rm CE}$ & -23.1 & -23.7 & 30 \\ 
97 & 2.0-2.5 & 0.5 & 300 & off & off & Tides & -23.4 & -24.0 & 43 \\ 
{\bf 98} & 2.0-2.5 & 0.5 & 300 & off & off & Twin Binaries & -22.7 & -23.4 & 15 \\ 
99 & 2.0-2.5 & 0.5 & 300 & off & off & Wind 1 & -23.0 & -23.7 & 29 \\ 
100 & 2.0-2.5 & 0.5 & 300 & off & off & Wind 2 & -23.8 & -24.3 & 49 \\ 
101 & 2.0-2.5 & 0.5 & 50 & off & off & V$_{\rm kick, ECS}$ = 50 & -31.7 & -33.2 & 93 \\ 
102 & 2.0-2.5 & 1.0 & 150 & off & off &  & -23.0 & -25.6 & 70 \\ 
103 & 2.0-2.5 & 1.0 & 50 & off & off &  & -23.3 & -50.0 & 110 \\ 
104 & 2.0-2.5 & 1.0 & 0 & off & off &  & -50.0 & -50.0 & 109 \\  
105 & 2.0-2.5 & 1.0 & 300 & off & off & V$_{\rm kick, ECS}$ = 60 & -50.0 & -50.0 & 108 \\ 
106 & 2.0-2.5 & 1.0 & 300 & off & off & V$_{\rm kick, ECS}$ = 18 & -50.0 & -50.0 & 106 \\ 
107 & 2.0-2.5 & 1.0 & 300 & off & off & V$_{\rm kick, ECS}$ = 6 & -50.0 & -50.0 & 107 \\ 
108 & 2.0-2.5 & 1.0 & 300 & off & off & V$_{\rm kick, ECS}$ = 0 & -50.0 & -50.0 & 105 \\ 
109 & 2.0-2.5 & 1.0 & 300 & on & off & F$_{\rm a}$ = 0.1 & -24.8 & -26.8 & 78 \\ 
110 & 2.0-2.5 & 1.0 & 300 & on & off & F$_{\rm a}$ = 1 & -23.1 & -50.0 & 104 \\ 
111 & 2.0-2.5 & 1.0 & 300 & on & off & $M_{\rm NS,max}$ & -23.3 & -50.0 & 103 \\ 
112 & 2.0-2.5 & 1.0 & 300 & on & off & $M_{\rm He,con}$ = 2 & -23.1 & -50.0 & 102 \\ 
113 & 2.0-2.5 & 1.0 & 300 & on & off & $\gamma_{\rm CE}$ & -22.5 & -24.3 & 48 \\ 
114 & 2.0-2.5 & 1.0 & 300 & on & off & Tides & -23.2 & -50.0 & 101 \\ 
115 & 2.0-2.5 & 1.0 & 300 & on & off & Twin Binaries & -22.9 & -25.1 & 63 \\ 
116 & 2.0-2.5 & 1.0 & 300 & on & off & Wind 1 & -23.2 & -25.6 & 69 \\ 
117 & 2.0-2.5 & 1.0 & 300 & on & off & Wind 2 & -23.2 & -25.7 & 71 \\ 
118 & 2.2-3.0 & 0.01 & 150 & off & off &  & -50.0 & -50.0 & 100 \\ 
119 & 2.2-3.0 & 0.05 & 150 & off & off &  & -50.0 & -50.0 & 99 \\ 
120 & 2.2-3.0 & 0.1 & 300 & off & off &  & -28.7 & -50.0 & 98 \\ 
121 & 2.2-3.0 & 0.2 & 300 & off & off &  & -25.8 & -28.2 & 87 \\ 
122 & 2.2-3.0 & 0.3 & 150 & off & off &  & -25.4 & -28.6 & 88 \\ 
123 & 2.2-3.0 & 0.3 & 300 & off & off &  & -25.0 & -29.0 & 89 \\ 
124 & 2.2-3.0 & 0.3 & 50 & off & off &  & -24.2 & -28.0 & 86 \\ 
125 & 2.2-3.0 & 0.5 & 150 & off & off &  & -25.8 & -27.9 & 85 \\ 
126 & 2.2-3.0 & 0.5 & 300 & off & off &  & -24.8 & -27.1 & 82 \\ 
127 & 2.2-3.0 & 0.5 & 50 & off & off &  & -24.4 & -27.4 & 84 \\ 
128 & 2.2-3.0 & 1.0 & 150 & off & off &  & -25.1 & -26.9 & 80 \\ 
129 & 2.2-3.0 & 1.0 & 300 & off & off &  & -25.1 & -26.6 & 77 \\ 
130 & 2.2-3.0 & 1.0 & 50 & off & off &  & -25.0 & -50.0 & 97 \\ 
131 & 2.2-3.0 & 1.0 & 300 & off & on &  & -23.5 & -26.0 & 72 \\ 
132 & 2.2-3.0 & 0.3 & 300 & off & on &  & -25.5 & -29.2 & 91 \\ 
133 & 2.2-3.0 & 1.0 & 300 & off & off & F$_{\rm a}$ = 0.1 & -28.0 & -29.2 & 90 \\ 
134 & 2.2-3.0 & 1.0 & 300 & off & off & F$_{\rm a}$ = 1 & -26.8 & -50.0 & 96 \\ 
135 & 2.2-3.0 & 1.0 & 300 & off & off & $\gamma_{\rm CE}$ & -24.5 & -27.3 & 83 \\ 
136 & 2.2-3.0 & 1.0 & 300 & off & off & Tides & -24.8 & -26.4 & 73 \\ 
137 & 2.5-2.7 & 1.0 & 300 & off & off &  & -23.2 & -24.4 & 53 \\ 
138 & 1.3-2.25 & 1.0 & 300 & off & off &  & -23.2 & -50.0 & 53 \\ 
139 & 1.66-3.24 & 1.0 & 300 & off & on &  & -24.2 & -50.0 & 154 \\ 
140 & 1.66-3.24 & 0.5 & 300 & off & on &  & -26.3 & -50.0 & 153 \\ 
141 & 1.66-3.24 & 0.01 & 150 & off & off &  & -50.0 & -50.0 & 152 \\ 
142 & 1.66-3.24 & 0.05 & 150 & off & off &  & -34.6 & -50.0 & 151 \\ 
143 & 1.66-3.24 & 0.1 & 150 & off & off &  & -30.5 & -50.0 & 150 \\ 
144 & 1.66-3.24 & 0.1 & 300 & off & off &  & -30.9 & -50.0 & 149 \\ 
145 & 1.66-3.24 & 0.3 & 150 & off & off &  & -25.6 & -32.3 & 92 \\ 
146 & 1.66-3.24 & 0.3 & 300 & off & off &  & -26.4 & -50.0 & 148 \\ 
147 & 1.66-3.24 & 0.3 & 50 & off & off &  & -26.8 & -50.0 & 147 \\ 
148 & 1.66-3.24 & 0.5 & 150 & off & off &  & -27.4 & -50.0 & 146 \\ 
149 & 1.66-3.24 & 0.5 & 300 & off & off &  & -25.9 & -50.0 & 145 \\ 
150 & 1.66-3.24 & 0.5 & 50 & off & off &  & -28.8 & -50.0 & 144 \\ 
151 & 1.66-3.24 & 1.0 & 150 & off & off &  & -31.4 & -35.5 & 94 \\ 
152 & 1.66-3.24 & 1.0 & 300 & off & off &  & -26.4 & -50.0 & 143 \\ 
153 & 1.66-3.24 & 1.0 & 50 & off & off &  & -31.7 & -35.6 & 95 \\ 
154 & 1.7-1.9 & 1.0 & 300 & off & off &  & -23.8 & -50.0 & 142 \\ 
155 & 1.8-2.1 & 1.0 & 300 & off & off &  & -23.5 & -50.0 & 141  
\enddata 

\tablenotetext{a}{The He core mass range within which a partially degenerate CO core forms. See \S \ref{ref_model}.} 
\tablenotetext{b}{The kick velocity applied to a NS born in an Fe core-collapse SN is drawn randomly from a Maxwellian distribution with this dispersion velocity.} 
\tablenotetext{c}{If on, NS accrete hypercritically in the CE.} 
\tablenotetext{d}{If on, HG stars are allowed to survive a CE.} 
\tablenotetext{e}{$\beta$: the specific angular momentum of matter; $M_{\rm NS,max}$: the maximum mass of a NS; $M_{\rm He,con}$: He stars below this mass develop convective envelopes; F$_{\rm a}$: The fraction of mass lost during stable mass transfer; $\gamma_{\rm CE}$: CE evolution is determined based on angular momentum conservation, not energy; Tides: Tidal dissipation is decreased by a factor of 5; Twin Binaries: The initial mass function for the secondary is chosen so that the mass ratio is closer to 1; Wind 1: He star winds are decreased by a factor of 4; Wind 2: H and He star winds are decreased by a factor of 4; V$_{\rm kick, ECS}$: Kick velocities for NS born in ECS are drawn from Maxwellian distributions with this dispersion velocity in km s$^{-1}$ instead of 1/10th that of Fe core-collapse SN.} 
\tablenotetext{f}{The rank of each model when ordered by $\Lambda_{\rm evol}$. Models with rank 1-20 are in bold.}
\end{deluxetable*}